\documentclass[conference]{IEEEtran}

\usepackage{cite}

\usepackage[dvipsnames]{xcolor}   %

\usepackage{graphicx}

\usepackage{amsmath}
\usepackage{amssymb}              %

\usepackage{amsthm}               %

\usepackage{array}
\usepackage{booktabs}
\usepackage{comment}            %
\usepackage{multirow}

\usepackage[caption=false,font=footnotesize]{subfig}

\usepackage{algorithm}
\usepackage{algpseudocode}

\usepackage[pdfauthor={},pdftitle={},pdfsubject={},pdfkeywords={},
            pdfcreator={},pdfproducer={}]{hyperref}

\algrenewcommand{\algorithmiccomment}[1]{\hfill\textcolor{blue}{// #1}}

\begin{document}

\title{Hearing the Whispers: Black-Box Membership Inference Attacks on Finetuned TTS Models}

\author{
\IEEEauthorblockN{
Kunlin Cai$^{\dagger,*}$,
Kaiyuan Zhang$^{\dagger,*}$,
Zihang Xiang$^{\dagger}$,
Jinghuai Zhang$^{\dagger}$,
Abeer Alwan$^{\dagger}$,
Fnu Suya$^{\ddagger}$, and
Yuan Tian$^{\dagger}$
}
\IEEEauthorblockA{
$^{\dagger}$University of California, Los Angeles
$^{\ddagger}$University of Tennessee, Knoxville\\
{\footnotesize $^{*}$Equal contribution.}
}
}

\maketitle

\begin{abstract} %

Text-to-Speech (TTS) foundation models are increasingly fine-tuned on private datasets to
synthesize highly personalized voices, introducing severe privacy risks by exposing both
biometric identities and sensitive speech content. Existing black-box membership inference
attacks (MIAs) follow a two-stage pipeline of query generation and representation
engineering, both of which face unique challenges when adapted to TTS. For query generation,
dual conditioning on synthesis text and reference speech creates a large and underexplored
query design space with no established criterion for identifying an effective query. For
representation engineering, the multi-level speech characteristics and temporal variability of
speech make low-level representations and direct comparisons inadequate for capturing
membership signals. To address these challenges, we present the first black-box MIA framework explicitly tailored to TTS
models at both the speaker and record levels. For query generation, we characterize the
feasible query space and establish two criteria, scorable extent and memorization elicitation,
for evaluating five representative queries, identifying recitation as the strongest. For
representation engineering, we obtain multi-level speech representations from high-fidelity
embedding models and temporally align the generated and target audio for fine-grained
comparison. Evaluations across three state-of-the-art TTS models (CosyVoice2, F5-TTS, and
XTTS-v2) fine-tuned on two benchmark datasets (VCTK and British Dialect) reveal severe privacy
leakage: speaker-level AUC remains above 0.80 and approaches 1.0 in the strongest settings,
while record-level AUC ranges from 0.80 to 0.90 and remains effective even in challenging scenarios where both members and non-members are of the same speakers. We further identify speech characteristics
associated with disproportionate vulnerability to memorization.
\end{abstract}


\IEEEpeerreviewmaketitle

\section{Introduction}

Modern Text-to-Speech (TTS) foundation models are routinely fine-tuned on private speech corpora to clone specific voices for applications such as AI-driven digital humans, audiobook narration, and voice agents~\cite{wang2023neural,shen2023naturalspeech}. Such fine-tuning can expose two forms of sensitive information: \textit{(i)} biometric identity, which can enable deepfakes and identity spoofing, and \textit{(ii)} the linguistic content of the training recordings, which may contain private conversations or medical consultations~\cite{backstrom2025privacy,tomashenko2022voiceprivacy,warren2024better}. As regulations such as GDPR~\cite{gdpr2016} and the EU AI Act~\cite{euaiact2024} increase scrutiny over the use of personal data in model training, auditing membership leakage from fine-tuned TTS models is therefore an important privacy problem.

Membership Inference Attacks (MIAs)~\cite{shokri2017membership, carlini2022membership} are the standard auditing tool for this question, but their effectiveness depends on tailoring the attack to the target modality and task~\cite{hu2022membershipsok, salem2023sok}. While MIAs have been extensively studied for vision~\cite{wu2022membership, pang2023black} and language~\cite{duan2024membership} models, audio work has focused on discriminative tasks such as ASR~\cite{shah2021evaluating} and speaker verification~\cite{chen2023slmia}. No prior work has built a black-box MIA for modern generative TTS, despite the fact that fine-tuning embeds targeted identity and linguistic traces directly into model parameters.

Existing black-box MIAs on generative models generally follow a two-stage pipeline: constructing queries that expose what the model has memorized from its training data, and engineering representations of the model responses to quantify the elicited behavior as a membership score~\cite{pang2023black,li2024towards}. However, adapting this design to TTS is challenging at both stages due to the dual-conditioning structure of TTS queries and the unique characteristics of speech outputs.

First, \textit{query generation}. Unlike prior generative models with a single conditioning
modality, modern black-box TTS models are jointly conditioned on both text and reference speech. This dual-conditioning creates a large and underexplored query design space, as the adversary must
determine what text to synthesize and what speech to provide as the reference. Within this space, it remains unknown what the candidate queries are, and there is no
criterion for judging which of them best elicits membership-dependent behavior. Query design
therefore has no principled basis to build on.

Second, \textit{representation engineering}: Even when a query exposes what the model has memorized from its training data, the resulting evidence is embedded in the synthesized speech, raising the question of how to extract and represent it as a measurable membership signal. This challenge stems from two properties specific to speech. \textit{(i) Multi-level speech characteristics.} Membership traces may manifest across speech characteristics at different levels, ranging from local spectral patterns to global speaking style. Such heterogeneous information cannot be adequately captured by a single low-level acoustic representation, such as Mel-spectrograms~\cite{stevens1937scale} or MFCCs~\cite{davis1980comparison}. \textit{(ii) Temporal variability.} Speech is a continuous, variable-length signal, and generations of the same content can differ in duration and pacing. This variability prevents direct application of the token-level matching~\cite{carlini2021extracting,shi2023detecting} and fixed-grid distance metrics~\cite{duan2023diffusion,pang2023black} used in prior MIAs.

To address these challenges, we present the first black-box MIA framework for fine-tuned TTS, operating at both speaker and record granularity. Our contributions are as follows:

\begin{itemize}
    \item We propose the first systematic membership inference study of modern generative TTS, formalizing the threat model at two granularities (speaker-level and record-level audits) and designing a black-box attack pipeline that resolves both obstacles above in a unified framework. This exposes a previously unexplored attack surface that arises specifically from fine-tuning on private voice data. (Section~\ref{sec:threat_model}).

    \item We present the first systematic study of query design for TTS MIA, which offers both a
    map of the query space and a principled way to navigate it. We characterize the feasible
    space along attacker knowledge and attack strategy, instantiating it into five representative
    queries. We further derive two criteria inspired by black-box MIAs
    to evaluate different queries for TTS MIA: \emph{scorable extent}, how much
    of the generated speech can be compared against the target, and \emph{memorization
    elicitation}, how strongly the query drives the model to reveal what it memorized from the
    target. Recitation, which queries the model with the target record under its training-time
    conditions, is the only candidate that fully satisfies both, and empirically performs strongest while other query strategies satisfy partially and have inferior performance. 
    (Sections~\ref{sec:query_generation}
    and~\ref{sec:ablation_query}). 

    \item We design two tailored representation extractors for continuous, variable-length speech: a speaker verification encoder that distills a global identity voiceprint for speaker-level audits, and multi-level WavLM~\cite{chen2022wavlm} embeddings that preserve frame-level memorization artifacts for record-level audits. To compare variable-length outputs, we align generated audio to the target time axis with a modified Dynamic Time Warping~\cite{sakoe2003dynamic} and aggregate the resulting non-linear similarity vectors through a lightweight LSTM scorer (Section~\ref{sec:representation_learning}).

    \item We conduct extensive evaluations across three state-of-the-art TTS models (CosyVoice2~\cite{du2024cosyvoice}, XTTS-v2~\cite{casanova2024xtts}, F5-TTS~\cite{chen2025f5}) fine-tuned on two benchmark datasets (VCTK~\cite{yamagishi2019cstr}, British Dialect~\cite{demirsahin2020open}). Speaker-level AUC is near perfect in the strongest settings and stays above 0.80 throughout; record-level AUC ranges from 0.80 to 0.90 and remains effective even when non-members are drawn from speakers already seen during fine-tuning. A finer analysis identifies which speech properties (low silence ratio, dense phoneme structure, rich high-frequency content) drive disproportionate vulnerability (Section \ref{sec:experiment}).
\end{itemize}

\section{Background}
\label{background}
This section covers the background of our work: the fine-tuning paradigm of modern TTS models (Section~\ref{background:tts}), the black-box membership inference attack (MIA) framework (Section~\ref{backgroud:MIA}), and the threat model for our MIAs on TTS (Section~\ref{sec:threat_model}).

\subsection{Text to Speech (TTS) Models}
\label{background:tts}
TTS synthesis converts written text into natural-sounding speech with the voice characteristics of a target speaker, and it serves as a core component in voice assistants, audiobook narration, and accessibility tools~\cite{wood2018does,tan2021survey, kathiria2024assistive}. Early TTS systems, such as Tacotron2~\cite{shen2018natural} and VITS~\cite{kim2021conditional}, typically synthesize speech for a closed set of pre-registered speakers. Modern TTS foundation models have moved beyond this limitation. By conditioning on a short reference utterance $s_{\text{ref}}$, these models can synthesize speech that matches the voice of an arbitrary speaker without dedicated speaker-specific fine-tuning. This capability has enabled broad applications in personalized voice generation, voice cloning, and content creation. Formally, given a synthesis text $t_{\text{syn}}$ and a reference utterance $s_{\text{ref}}$, it generates a waveform $s_{\text{gen}}$ as follows:
\begin{equation}
s_{\text{gen}} = \mathcal{F}(t_{\text{syn}}, s_{\text{ref}}; \theta)
\end{equation}
where $\mathcal{F}$ denotes a generative model parameterized by $\theta$ that is conditioned on $s_{\text{ref}}$ to preserve the acoustic characteristics of the target speaker.

Modern TTS architectures can be broadly categorized as autoregressive (AR), non-autoregressive (NAR), and hybrid models~\cite{xie2025towards, azzuni2025voice}. AR models, such as VALL-E~\cite{wang2023neural}, XTTS-v2~\cite{casanova2024xtts}, and Qwen3-TTS-12Hz~\cite{hu2026qwen3}, generate speech token by token. They are effective at modeling fine-grained prosody and natural expressiveness, but often incur higher inference latency. Non-autoregressive models usually employ diffusion or flow-matching decoders to directly map text to mel-spectrograms, with representative examples including Voicebox~\cite{le2023voicebox}, E2-TTS~\cite{eskimez2024e2}, and F5-TTS~\cite{chen2025f5}. These models enable faster inference and stronger duration control, although they may reduce stylistic diversity in some settings. Hybrid models, such as CosyVoice~\cite{du2024cosyvoice1, du2024cosyvoice}, Chatterbox TTS~\cite{chatterboxtts2025}, and Qwen3-TTS-25Hz~\cite{hu2026qwen3}, combine autoregressive and non-autoregressive components to balance expressiveness and efficiency. In this work, we evaluate the privacy vulnerabilities of fine-tuned TTS systems using three representative and widely-used models.

\subsection{Membership Inference Attacks}
\label{backgroud:MIA}
Membership Inference Attack (MIA) is a prominent framework for auditing the privacy leakage of machine learning models. Its primary objective is to determine whether a specific data record $x$ was included in the target model's training dataset. In practical deployment scenarios, adversaries typically operate under a \emph{black-box} threat model. Under this setting, the adversary cannot observe the model's internal parameters or gradients, and must instead rely entirely on the observable outputs to perform inference.

Recent studies on conditional generative models often formulate black-box MIAs as a score-based decision process~\cite{pang2023black}. Rather than querying the target model $\mathcal{F}$ directly with the exact training record $x$, the adversary crafts an attack query $q$ to prompt the model. The adversary then derives a scalar \emph{membership score} $\mathcal{M}(x)$ to quantify the likelihood of $x$ being in the training set. Formally, this score is computed via a feature extraction function $\phi(\cdot)$:
\begin{equation}
    \mathcal{M}(x) = \phi\big(x, \mathcal{F}(q)\big)
\end{equation}
where $\mathcal{F}(q)$ represents the model's generated output in response to the query. The function $\phi(\cdot)$ serves to capture the similarity, reconstruction error, or structural representation gap between the original target sample $x$ and the generated output. The final membership decision is made by comparing $\mathcal{M}(x)$ against a predefined threshold $\tau$. The effectiveness of the attack critically depends on designing a robust $\phi(\cdot)$ that captures the model's disparate behavior on seen versus unseen data.

Applying this framework to TTS is non-trivial, because speech breaks both of the components above. Constructing the query $q$ is no longer straightforward: the input space is dual-modality (a synthesis text plus a reference audio), and it is unclear which combination of the two most effectively triggers memorization. Instantiating $\phi(\cdot)$ is difficult for two further reasons. First, TTS outputs are continuous waveforms in which linguistic content, speaker identity, and prosody are heavily entangled, so subtle membership traces are hard to isolate from the dominant generic content that any competent model would produce. Second, the generated waveforms are variable-length with stochastic timing, so the target record and the model's output cannot be compared via point-to-point distance metrics even once informative features are available.

\subsection{Threat Model}
\label{sec:threat_model}
\noindent\textbf{Target Scenario.} We focus on the vulnerability of TTS models during the fine-tuning phase. Generative TTS models are frequently fine-tuned on private, proprietary, or copyrighted voice datasets ($\mathcal{D}_{\text{train}}$) to adapt them to specific domains, languages, or premium speaker identities. The target system is a fine-tuned TTS model $\mathcal{F}$, and the dataset $\mathcal{D}_{\text{train}}$ is kept confidential by the model provider.

\noindent\textbf{Adversarial Goal.} The adversary acts as an external auditor (e.g., a privacy regulator, a copyright holder, or a security researcher) aiming to assess the privacy leakage of $\mathcal{F}$. To achieve this, their primary objective is to conduct a Membership Inference Attack that determines whether a target record $x$ or a target speaker identity was included in $\mathcal{D}_{\text{train}}$. Functionally, this attack is framed as a strict binary classification task (member versus non-member) aimed at verifying unauthorized data usage or compliance violations.

\noindent\textbf{Adversary's Capabilities.}
We assume a practical \emph{black-box} setting.
The adversary has no knowledge of the model's internal weights, gradients, or fine-tuning
data. Specifically, the adversary can submit multiple attack queries $q = (t_{syn}, s_{\text{ref}})$,
consisting of a text prompt and a reference audio, and observe the generated continuous
acoustic waveform $s_{\text{gen}} = \mathcal{F}(q)$. The adversary cannot access internal
confidence scores, token-level probabilities, or intermediate representations. However,
consistent with realistic auditing scenarios, we assume the adversary has access to standard
computational resources and publicly available auxiliary tools. Specifically, the adversary
can leverage open-source, pre-trained embedding models as surrogate feature extractors to
process and analyze the generated acoustic outputs post-hoc. Following standard practice in
membership inference~\cite{shokri2017membership,carlini2022membership}, the record-level
attack additionally assumes the adversary knows which publicly released base checkpoint the
deployed system is built on and can fine-tune its own shadow copy on disjoint auxiliary
data; the speaker-level attack requires no such assumption.

\noindent\textbf{Auditing Granularities.} Auditing TTS relies on conditioning variables that often differ from the training audio. We formalize this threat model into two distinct levels, operating under different assumptions of adversarial knowledge (summarized in Table~\ref{tab:threat_model}):

\textbf{(1) Speaker-Level MIA.} The adversary aims to determine if \emph{any} data belonging to a target speaker $\text{spk}_{\text{target}}$ is in $\mathcal{D}_{\text{train}}$. Here, we assume the adversary possesses $n$ distinct audio clips of $\text{spk}_{\text{target}}$, without knowing if they overlap with the training set. This represents a realistic scenario where a public figure suspects unauthorized voice cloning and uses public recordings as queries to detect the memorization of their general acoustic profile.

\textbf{(2) Record-Level MIA.} The adversary aims to determine if a \emph{specific} text-audio record $x = (t_{\text{target}}, s_{\text{target}})$ is in $\mathcal{D}_{\text{train}}$. We assume the adversary possesses the exact target record $x$. This is a more challenging, fine-grained task with assumptions aligning with standard MIA, representing a scenario where a user audits the unauthorized use of a specific private recording (e.g., a leaked voice message) during fine-tuning.

\begin{table}[htbp]
\small
\centering
\caption{Summary of Threat Model Assumptions at Different Auditing Granularities}
\label{tab:threat_model}
\begin{tabular}{@{}p{0.20\linewidth}p{0.35\linewidth}p{0.37\linewidth}@{}}
\toprule
\textbf{Attack} & \textbf{Adversary Objective} & \textbf{Adversarial Knowledge} \\
\midrule
\addlinespace
\textbf{Speaker-Level MIA} & Detect if a target speaker $\text{spk}_{\text{target}}$ $\in \mathcal{D}_{\text{train}}$. & A set of $n$ reference utterances from $\text{spk}_{\text{target}}$. \\
\addlinespace
\midrule
\addlinespace
\textbf{Record-Level MIA} & Detect if the target record $x \in \mathcal{D}_{\text{train}}$. & A target text-audio pair $x = (t_{\text{target}}, s_{\text{target}})$. \\
\addlinespace
\bottomrule
\end{tabular}
\end{table}

\section{Methodology}\label{sec:method}

In this section, we present our black-box MIA framework tailored for fine-tuned TTS models. On standard generative models, a black-box MIA typically proceeds in two stages: query generation, which probes the model with inputs derived from the target record, and representation engineering, which extracts features from the generated output and compares them with the target to obtain a membership score. Porting this pipeline to TTS raises a challenge at each stage, and neither has been examined by prior work.

The first challenge concerns query generation. A TTS model is jointly conditioned on a synthesis text and a reference speech. This creates a far broader query space than the single-condition settings considered in prior MIAs. More fundamentally, since query design under this dual-conditioning setting has not been studied, there is neither an established query to adopt nor a standard for judging whether a candidate query is strong. We therefore characterize the space of feasible queries and instantiate it into five representative queries. To judge them, we abstract two evaluation criteria, scorable extent and memorization elicitation, from the reconstruction strategy shared by black-box MIAs; recitation ranks first under both criteria and serves as the query for our attack (Section~\ref{sec:query_generation}).

The second challenge concerns representation engineering. Membership traces embed in speech characteristics at multiple levels, which low-level spectral descriptors such as Mel-spectrograms~\cite{stevens1937scale} and MFCCs~\cite{davis1980comparison} cannot capture. In addition, TTS models generate continuous, variable-length waveforms with inherent temporal stochasticity, so the point-to-point comparison available for discrete tokens or fixed-size images does not apply, and the generated speech must be aligned with the target without destroying the fragile membership artifacts. We address both issues with tailored representation extraction and alignment strategies (Section~\ref{sec:representation_learning}).

Algorithm~\ref{alg:mia_tts} in the Appendix summarizes the framework: an $m$-query sampling strategy built on a reconstruction-based query generator $G_Q$ and a representation pipeline $\phi(\cdot,\cdot)$ that extracts, aligns, and scores the generated speech.

\subsection{Query Generation ($G_Q$)}
\label{sec:query_generation}

The first stage of our attack centers on how to query the target model. The importance of query
design for effective membership inference has been well established in prior
studies~\cite{wu2022membership,pang2023black,zhai2024membership,carlini2023extracting}. However,
unlike attacks on language or diffusion models, where a query typically consists of a single
text prompt, a TTS query $q=(t_{\text{syn}}, s_{\text{ref}})$ contains two coupled conditions: the
synthesis text and the reference speech. Each condition may exactly match the corresponding
component of the target record $x^{*}=(t^{*}, s^{*})$, cover only part of it, be a perturbed
version of it, or differ from it entirely. We therefore characterize the resulting query space
(§\ref{sec:query_space}), establish principles for evaluating queries for TTS MIA
(§\ref{sec:design_principle}), and identify recitation as the candidate.

\subsubsection{What Queries are available for TTS MIAs?}
\label{sec:query_space}
The coupled conditioning of TTS gives rise to a broader space of possible queries, making it unclear which query most effectively reveals membership. This question has not been systematically examined in prior work, motivating a systematic characterization of the query space. To this end, we follow established threat-modeling practices in prior privacy and membership inference studies~\cite{salem2023sok,hu2022membershipsok,nasr2021adversary}. Specifically, we consider variations in both attacker knowledge, such as whether the available information is partial or degraded, and attack strategy, such as prefix continuation, which follows the conditioning structure of autoregressive training. Based on these dimensions, we identify five representative queries, as summarized in Table~\ref{tab:query_space} and formally defined below.

\begin{table}[t]
\centering
\footnotesize
\setlength{\tabcolsep}{4pt}
\caption{Representative queries for TTS MIA. $x^{*}=(t^{*},s^{*})$ is the target record; $\rho\in(0,1)$ is a coverage fraction; $w_{\rho}(\cdot)$ extracts a window; $\mathcal{T}_{\sigma}$ and $\mathcal{P}_{\delta}$ denote acoustic and textual perturbations with strength $\sigma$ and $\delta$.}
\label{tab:query_space}
\begin{tabular}{@{}llll@{}}
\toprule
Query & $q=(t_{\text{syn}}, s_{\text{ref}})$ & Knowledge & Strategy \\
\midrule
Recitation         & $(t^{*},\, s^{*})$                            & full record      & reproduction \\
Continuation       & $(t^{*}_{\text{suf}},\, s^{*}_{\text{pre}})$  & full record      & prefix completion \\
Partial reference  & $(t^{*}_{\rho},\, w_{\rho}(s^{*}))$           & speech clip      & reproduction \\
Audio perturbation & $(t^{*},\, \mathcal{T}_{\sigma}(s^{*}))$      & degraded speech  & reproduction \\
Text perturbation  & $(\mathcal{P}_{\delta}(t^{*}),\, s^{*})$      & noisy transcript & reproduction \\
\bottomrule
\end{tabular}
\end{table}

\noindent\textbf{Recitation.} $q_{\text{rec}}=(t^{*},\,s^{*})$: an adversary with access to the full record asks the model to reproduce the record as is, using $s^{*}$ as the reference speech and $t^{*}$ as the synthesis text.

\noindent\textbf{Continuation.} $q_{\text{con}}=(t^{*}_{\text{suf}},\,s^{*}_{\text{pre}})$, where $s^{*}=s^{*}_{\text{pre}}\,\|\,s^{*}_{\text{suf}}$ and $t^{*}=t^{*}_{\text{pre}}\,\|\,t^{*}_{\text{suf}}$ partition the record at a fraction $\rho\in(0,1)$ ($\|$ denotes concatenation): the adversary has access to the full record but deliberately withholds the suffix and asks the model to complete it given the speech prefix $s^{*}_{\text{pre}}$. This design follows the prefix completion strategy widely used to probe memorization in black-box LLMs~\cite{carlini2021extracting,carlini2023extracting,ippolito2022preventing}, under which a memorized sequence is expected to be continued more faithfully than an unseen sequence.

\noindent\textbf{Partial reference.} $q_{\text{par}}=(t^{*}_{\rho},\,w_{\rho}(s^{*}))$, where $w_{\rho}(s^{*})$ denotes a window at an arbitrary position that covers a fraction $\rho$ of $s^{*}$, and $t^{*}_{\rho}$ denotes the corresponding transcript: the adversary has access to only a clip of the target speech and asks the model to reproduce that clip.

\noindent\textbf{Audio perturbation.} $q_{\text{aud}}=(t^{*},\,\mathcal{T}_{\sigma}(s^{*}))$, where $\mathcal{T}_{\sigma}$ denotes an acoustic perturbation of strength $\sigma$, such as additive noise: the adversary has access to only a degraded copy of the target speech and asks the model to reproduce $t^{*}$ using the degraded speech as the reference.

\noindent\textbf{Text perturbation.} $q_{\text{txt}}=(\mathcal{P}_{\delta}(t^{*}),\,s^{*})$, where $\mathcal{P}_{\delta}$ replaces a fraction $\delta$ of the words with near-homophonic alternatives: the adversary has access to the clean target speech but only an imperfect transcript, such as one produced by ASR, and asks the model to synthesize the perturbed transcript using $s^{*}$ as the reference speech.

\subsubsection{Which Query Design Is Most Effective for MIA? }
\label{sec:design_principle}

Given the query candidates above, the key question is which query produces the largest
membership-dependent reconstruction gap. Since black-box MIA against modern TTS models has not been
established, we draw on black-box MIAs against other generative models, which commonly prompt the
model to reconstruct a target record and measure the similarity between the reconstruction and the
target~\cite{pang2023black,li2024towards}. This follows from the connection between membership and
training loss~\cite{yeom2018privacy}: training loss is inaccessible through a black-box API, but
reconstruction quality is an observable proxy for how well the model fits a particular
record~\cite{chen2020ganleaks}. The intuition transfers to TTS because modern TTS models are
themselves trained through reconstruction, learning to reproduce an utterance from its transcript
and a reference segment taken from the same
utterance~\cite{casanova2024xtts,du2024cosyvoice,chen2025f5,le2023voicebox,zhu2026omnivoice}. We
therefore build our attack around reconstruction similarity: reproducing the target speaker's
identity at the speaker level, and the target recording's local acoustic realization at the record
level.

High reconstruction similarity alone, however, does not constitute a membership signal. A query is
useful only when it causes member outputs to match the target more closely than non-member outputs,
widening the gap between their reconstruction scores. We therefore evaluate each query according to
two criteria: (C1) \emph{scorable extent}, how much membership-relevant evidence in the generated
speech can be meaningfully compared with the target, primarily determined by the synthesis text
$t_{\mathrm{syn}}$; and (C2) \emph{memorization elicitation}, how strongly the query causes the
model to express target-specific information learned during fine-tuning, primarily determined by
the reference speech $s_{\mathrm{ref}}$. We discuss both below and test the resulting predictions in
Section~\ref{sec:ablation_query}.

\noindent\textbf{C1: A strong query should maximize the scorable extent.}
For record-level MIA, the generated and target speech must share linguistic content before their acoustic realizations can be meaningfully compared. Otherwise, their distance conflates differences in speaker characteristics, prosody, and fine-grained pronunciation with differences caused simply by speaking different words. The synthesis text therefore determines how much of the target record is scorable. When $t_{\mathrm{syn}}=t^{*}$, the entire generated utterance corresponds to $s^{*}$. When $t_{\mathrm{syn}}$ covers only part of $t^{*}$, only the corresponding portion of $s^{*}$ can be evaluated. When $t_{\mathrm{syn}}$ modifies part of $t^{*}$, the altered words have no content-matched counterpart in $s^{*}$ and cannot provide reliable reconstruction evidence.

For speaker-level MIA, exact lexical correspondence matters less because the attack compares
identity-focused embeddings that suppress linguistic variation; scorable extent instead refers to
the amount of speech available for extracting a speaker representation. The more distinct
linguistic content the query induces the model to produce, the broader the phonetic coverage of the
generated speech and the more identity evidence there is to compare against the target, whereas a
short suffix or window supplies less. Section~\ref{sec:spk_nutt_ablation} supports this empirically.

Under C1, the queries in Table~\ref{tab:query_space} consequently differ across the two auditing granularities. For record-level MIA, continuation and partial reference generate only a suffix or a window of $t^{*}$, so only part of the target record is scorable. Text perturbation generates a full-length utterance, but the fraction $\delta$ of replaced words has no content-matched counterpart in $s^{*}$. Recitation and audio perturbation use the complete, unaltered $t^{*}$ and therefore preserve the full scorable extent. For speaker-level MIA, recitation, audio perturbation, and text perturbation all produce full-length speech from which identity can be extracted, while continuation and partial reference provide shorter and potentially less stable identity evidence.

\noindent\textbf{C2: A strong query should elicit memorized target details.}
Given sufficient scorable evidence, the member--non-member gap depends on whether the query preferentially exposes information learned from the target during fine-tuning. For a member, the model parameters may encode target-specific information that is not expressed under arbitrary conditioning. At the speaker level, this information includes detailed identity related characteristics such as habitual timbre detail and accent; at the record level, it includes details such as local pitch contours, pause positions and harmonic patterns. Providing a complete and faithful target reference is more likely to cue these learned details~\cite{carlini2022quantifying,carlini2023extracting,ren2022revisiting}.

A non-member model receives the same reference speech and can therefore transfer general properties such as speaker identity and overall speaking style. However, it has not learned the association between $t^{*}$ and the exact recorded realization in $s^{*}$. Even when the reference itself contains fine-grained record-specific details, the non-member model need not reproduce those details and may instead generate a more generic realization of the same content~\cite{zhou2022content}. Appendix~\ref{app:recon_example} illustrates this difference: under the same recitation query, the member reconstruction more closely matches the target's local harmonic and formant patterns, whereas the non-member reconstruction preserves the overall content and identity but misses these fine details.

A more complete reference can therefore raise reconstruction similarity for both members and non-members without affecting them equally. If the member output moves further toward the exact recorded realization, the member--non-member score gap increases rather than merely shifting both distributions upward. Appendix~\ref{app:ref_length} verifies this effect in a controlled experiment that fixes the synthesis text and varies only reference completeness. Supplying the full reference raises the scores of both populations but roughly doubles the gap between their means.

Under C2, continuation and partial reference provide only a segment of $s^{*}$, reducing the target information available to cue learned details. Audio perturbation provides the full reference in degraded form and is therefore expected to offer a weaker elicitation cue than the clean target speech. Recitation and text perturbation both provide the complete, unaltered $s^{*}$ and thus preserve the strongest conditioning available.

Combining C1 and C2 predicts recitation as the strongest overall query. At the record level, it is the only candidate that simultaneously provides the complete target transcript and the complete, clean target reference. Text perturbation preserves strong memorization elicitation but loses scorable extent over the replaced words; it should therefore remain competitive for speaker-level MIA, where exact linguistic correspondence is less important, while suffering a larger loss at the record level. Audio perturbation preserves the full scorable extent but weakens memorization elicitation, whereas continuation and partial reference restrict both criteria. Section~\ref{sec:ablation_query} evaluates these predictions empirically.

\subsection{Representation Engineering ($\phi$)}
\label{sec:representation_learning}
In this section, we detail the design of our representation engineering module ($\phi$). To enable successful MIAs, we address two fundamental challenges: first, we introduce tailored representation extractors to capture highly discriminative membership signals (Section~\ref{sec:extraction_method}); second, we design representation alignment techniques to resolve the inherent temporal misalignment of continuous audio (Section~\ref{sec:alignment_method}).

\subsubsection{Representation Extraction}
\label{sec:extraction_method}
To overcome the limitations of low-level acoustic representations (e.g., Mel-spectrograms~\cite{stevens1937scale} or MFCCs~\cite{davis1980comparison}) discussed earlier, we design tailored representation extractors aligned with the attack objectives: distilling stable global identity vectors via an speaker verification (SV) model for speaker-level MIA, and preserving prosodic artifacts via a self-supervised model for record-level MIA.

\noindent\textbf{Representation Extraction for Speaker-Level MIA.} Speaker-level MIA fundamentally relies on detecting a speaker's unique identity. This requires extracting a highly discriminative "identity voiceprint" from audio signals. Pre-trained SV models are well suited for this task, as they are optimized to condense high-dimensional audio into an identity-focused latent space~\cite{snyder2018x, DesplanquesTD20}. Accordingly, our evaluation function $\phi_{\text{spk}}(\cdot)$ relies on a WavLM + ECAPA - TDNN SV encoder~\cite{chen2022large} ($\mathcal{E}_{\text{SV}}$). Extracting features through this specialized architecture provides a robust representation for MIA.

Because SV systems are known to perform poorly on short audio segments~\cite{liu2020text}, averaging embeddings from individual clips fails to capture stable speaker traits. To address this limitation, we leverage the adversary’s access to $n$ target samples as defined in our threat model. Specifically, we temporally concatenate all $n$ audio sequences generated by querying the target model with these samples before extracting features.
This allows the SV model to process the entire sequence jointly. Specifically, we compute the fixed-dimensional global speaker embeddings, $\mathbf{e}_{\text{target}}$ and $\mathbf{e}_{\text{gen}}$, by passing the concatenated queries through the SV encoder $\mathcal{E}_{\text{SV}}$:
\begin{equation}
\begin{aligned}
    \mathbf{e}_{\text{target}} &= \mathcal{E}_{\text{SV}}(\text{Concat}(s_1, \dots, s_n)) \in \mathbb{R}^d, \\
    \mathbf{e}_{\text{gen}} &= \mathcal{E}_{\text{SV}}(\text{Concat}(s_{\text{gen}, 1}, \dots, s_{\text{gen}, n})) \in \mathbb{R}^d
\end{aligned}
\end{equation}
where $d$ denotes the embedding dimensionality. By translating the variable-length audio into these stable, fixed-dimensional vectors, we establish a reliable basis for computing the membership score in subsequent steps.

\noindent\textbf{Representation Extraction for Record-Level MIA.}
Detecting record-level memorization is more challenging than speaker attack. It requires capturing fine-grained, sequence-specific artifacts (e.g., exact phonetic alignment and micro-prosodic inflections) rather than a global identity invariant like speaker identity. 
Conversely, representations from SV models are also insufficient for this task. Because these models are explicitly optimized to discard localized linguistic and prosodic variations to isolate an utterance-independent voiceprint, applying them at the record level inadvertently removes the precise trajectory artifacts we aim to detect.

To resolve this representation gap, $\phi_{\text{rec}}(\cdot)$ evaluates trajectory reconstruction with a self-supervised WavLM~\cite{chen2022wavlm} encoder. The suitability of WavLM for this task is twofold. First, its masked speech prediction objective forces the transformer backbone to model local dependencies frame by frame, preserving the continuous micro-prosodic trajectory characteristic of exact memorization. Second, WavLM encodes a structured feature hierarchy across its $L=24$ transformer layers, with different layers capturing different types of information, as shown in the original paper. By extracting unpooled hidden states, we avoid the destructive abstraction of top-layer pooling and obtain robust multi-level representations that better capture membership signals. Formally, for an input utterance $x \in {\text{target}, \text{gen}}$, we construct the feature sequence as:
\begin{equation}
\mathbf{H}_{x} = \phi_{\text{rec}}(x) \in \mathbb{R}^{K_x \times |L_{\text{sel}}| \times d}
\end{equation}
where $K_x$ is the temporal frame length, $L_{\text{sel}}$ is the set of selected layers, and $d$ is the hidden state dimensionality ($d=1024$ for WavLM). Specifically, we denote $\mathbf{h}_{x, i}^{(l)} \in \mathbb{R}^d$ as the specific embedding vector extracted at temporal frame $i$ from layer $l$ (where $l \in L_{\text{sel}}$).

\subsubsection{Representation Temporal Alignment}
\label{sec:alignment_method}
In prior MIAs against text or image models, generated outputs are directly comparable. In contrast, TTS generates continuous, variable-length waveforms with inherently fluctuating durations and speaking rates, making naive frame-by-frame comparisons invalid. To resolve this temporal misalignment, we introduce distinct alignment strategies tailored to each of our proposed attacks. Specifically, for speaker-level MIA, our feature extractor collapse variable-length audio into fixed-dimensional global vectors for direct comparison. For record-level MIA, where fine-grained trajectory artifacts must be preserved, we design a modified Dynamic Time Warping (DTW)~\cite{sakoe2003dynamic} mechanism to dynamically align the frame-level sequences.

\noindent\textbf{Representation Alignment for Speaker-level MIA.} 
As previously mentioned, our feature extraction process yields fixed-dimensional vectors ($\mathbf{e}_{\text{target}}$ and $\mathbf{e}_{\text{gen}}$) that explicitly represent the speaker's identity. This fixed dimensionality allows us to directly compare the target and generated audio mathematically. More importantly, because SV models are trained to separate speaker identities based on the angle between their feature vectors, cosine similarity naturally serves as the optimal metric to measure their overlap. Specifically, our evaluation function $\phi_{\text{spk}}$ quantifies this acoustic representation gap as:
\begin{equation}
    v = \phi_{\text{spk}}(s_{\text{target}}^{\text{concat}}, s_{\text{gen}}^{\text{concat}}) = \frac{\mathbf{e}_{\text{target}} \cdot \mathbf{e}_{\text{gen}}}{\|\mathbf{e}_{\text{target}}\|_2 \|\mathbf{e}_{\text{gen}}\|_2}
\end{equation}
Because our prior temporal concatenation step effectively stabilizes these features, the score $v$ serves as a highly reliable measure of acoustic similarity and yields a discriminative membership indication to drive our threshold-based decision mechanism ($\tau$).

\begin{figure}[tbp]
    \centering

    \subfloat[Target speech.]{%
        \includegraphics[height=0.17\textheight,width=0.32\linewidth,keepaspectratio]{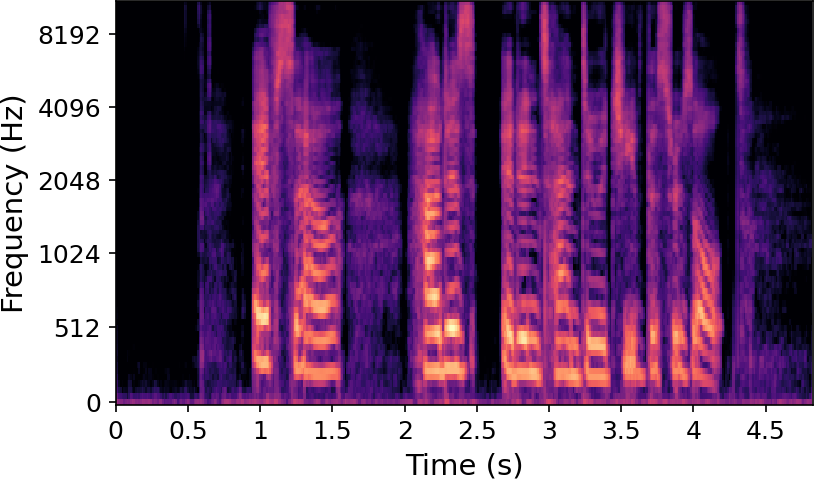}%
        \label{fig:rewrite_dtw_ref}}
    \hfill
    \subfloat[Generated speech.]{%
        \includegraphics[height=0.17\textheight,width=0.32\linewidth,keepaspectratio]{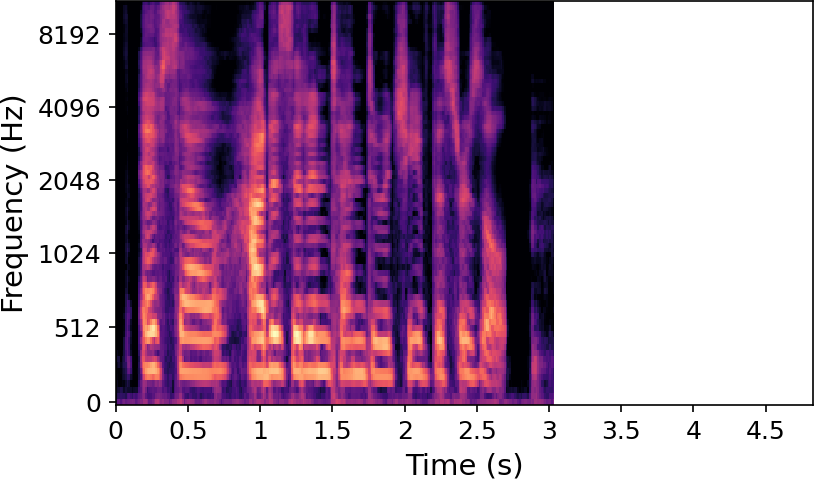}%
        \label{fig:rewrite_dtw_gen}}
    \hfill
    \subfloat[Aligned speech.]{%
        \includegraphics[height=0.17\textheight,width=0.32\linewidth,keepaspectratio]{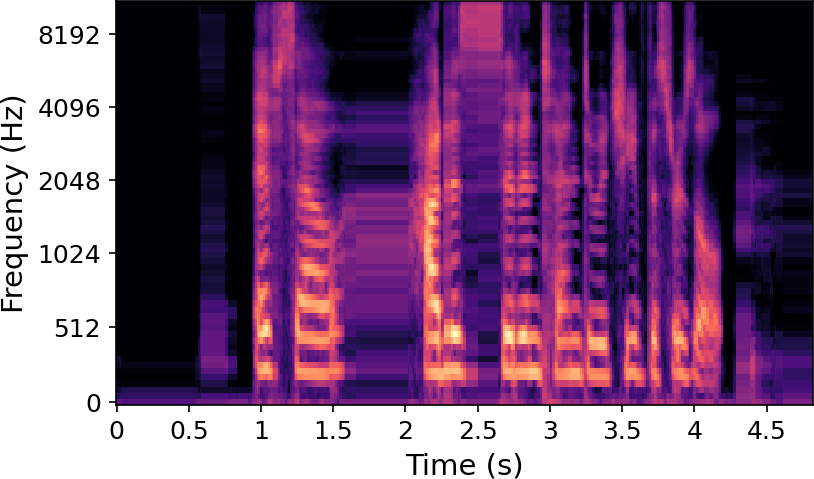}%
        \label{fig:rewrite_dtw_aligned}}

    \caption{DTW-based temporal alignment for record-level MIA. Even with matched transcriptions, the target speech (a) and the generated speech (b) differ in phoneme duration and prosodic timing. To enable fine-grained comparison, DTW warps the generated speech to produce (c), where corresponding spectral patterns are temporally aligned with the target speech (a).}
    \label{fig:rewrite_dtw_motivation}
\end{figure}

\noindent\textbf{Representation Alignment for Record-level MIA.} While the unpooled WavLM hidden states provide the frame-level resolution required to detect record-level memorization, the extracted feature of target record ($\mathbf{H}_{\text{target}}$) and generated results ($\mathbf{H}_{\text{gen}}$) cannot be directly compared frame by frame. As illustrated in Figure~\ref{fig:rewrite_dtw_motivation}, even under identical transcription, TTS models may synthesize the same phonetic sequence with different durations, speaking rates, and local prosodic timing. To obtain fine-grained temporal correspondence and enable consistent comparison across multiple queries, we align $\mathbf{H}_{\text{gen}}$ to $\mathbf{H}_{\text{target}}$ using a modified DTW, which warps the generated features to the time axis of the target utterance.

For each selected layer $l \in L_{\text{sel}}$ \emph{independently}, we compute a per-layer optimal warping path using cosine distance:
\begin{equation}
    \pi^{(l)\star}
    =
    \arg\min_{\pi \in \Pi}
    \sum_{(i,j)\in \pi}
    \left(1 -
    \frac{\hat{\mathbf{h}}_{\text{target}, i}^{(l)} \cdot \hat{\mathbf{h}}_{\text{gen}, j}^{(l)}}
    {\|\hat{\mathbf{h}}_{\text{target}, i}^{(l)}\|_2 \|\hat{\mathbf{h}}_{\text{gen}, j}^{(l)}\|_2}
    \right),
\end{equation}

where $\hat{\mathbf{h}}$ denotes the L2-normalized hidden state and $\Pi$ denotes the set of valid monotonic DTW paths. Unlike standard DTW, which extracts a similarity sequence along the warping path (producing a variable-length output), our modified DTW uses the path to \emph{warp the generated features onto the target time axis}. Specifically, for each target frame $i=1,\dots,K_{\text{target}}$, let $J^{(l)}_{i}=\{j : (i,j)\in\pi^{(l)\star}\}$ be the generated frames that $\pi^{(l)\star}$ maps to $i$; we average them:
\begin{equation}
    \mathbf{h}_{\text{warped}, i}^{(l)}
    =
    \frac{1}{|J^{(l)}_{i}|}
    \sum_{j \in J^{(l)}_{i}}
    \mathbf{h}_{\text{gen}, j}^{(l)}.
\end{equation}
The layer-wise cosine similarity between the target and the warped generated features at frame $i$ is then
\begin{equation}
    s^{(l)}_{i}
    =
    \frac{\mathbf{h}_{\text{target}, i}^{(l)} \cdot \mathbf{h}_{\text{warped}, i}^{(l)}}
    {\|\mathbf{h}_{\text{target}, i}^{(l)}\|_2\, \|\mathbf{h}_{\text{warped}, i}^{(l)}\|_2}.
\end{equation}
This produces a multivariate temporal similarity sequence $\mathbf{S} \in \mathbb{R}^{K_{\text{target}} \times |L_{\text{sel}}|}$, where $K_{\text{target}}$ is the length of the target utterance. By anchoring the output to the target time axis, this warping strategy ensures that the downstream LSTM receives a fixed-length sequence directly aligned with the target acoustic trajectory, eliminating variable-length outputs across utterances and queries.

Having extracted the temporally aligned similarity sequence $\mathbf{S}$, our final objective is to distill this two-dimensional representation into a single record-level score. However, reducing this sequence requires capturing temporal dependencies. Simple frame-wise averaging is suboptimal, as consecutive frames of high hyperspherical similarity provide a substantially stronger memorization signal than isolated matches. To capture this structural continuity, we employ a lightweight Long Short-Term Memory (LSTM) network as the attack classifier that aggregates the sequence over time and WavLM layers. Given the similarity sequence $\mathbf{S} = [\mathbf{s}_1, \mathbf{s}_2, \dots, \mathbf{s}_{K_{\text{target}}}]$, the temporal aggregation to a final score $v$ is formulated as:
\begin{equation}
\mathbf{m}_k = \text{LSTM}(\mathbf{s}_k, \mathbf{m}_{k-1}), \quad v = \sigma(\mathbf{W}^\top \mathbf{m}_{K_{\text{target}}} + b)
\end{equation}
where $\mathbf{m}_k$ represents the final hidden state encompassing the entire sequence context, and $\sigma$ denotes the projection mapping. To train the attack classifier, we fine-tune a shadow model of the same architecture on a disjoint speaker set with the same recipe (Section~\ref{sec:experiment_setup}), run the same query and representation pipeline against it, and use the resulting similarity sequences, whose membership labels are known, as training data. The trained classifier is then applied to the sequences obtained from the victim model. By evaluating continuous structural alignment rather than discrete frame similarities, the attack classifier robustly quantifies record-level memorization, yielding a high-fidelity scalar for the membership score $\mathcal{M}(x)$.

\section{Experiment}
\label{sec:experiment}

In the following, we illustrate our experimental setup in Section~\ref{sec:experiment_setup}. We empirically evaluate the proposed attack suite to answer the following research questions: (1) What is the performance of the proposed membership inference attacks against TTS models at both the speaker and record levels? (Section~\ref{sec:attack_performance}) (2) How do our attack designs including query, feature and alignment improve attack performance? (Sections~\ref{sec:ablation_query} to ~\ref{sec:alignment}) (3) How do the attack parameters affect attack performance? (Sections~\ref{sec:model_component_comparison} to ~\ref{sec:query_budget}) (4) What speech data characteristics correlate with membership inference vulnerability? (Section~\ref{sec:speech_factors}).

\subsection{Experiment Setup}
\label{sec:experiment_setup}

\noindent\textbf{Models.} We conduct experiments using the official implementations of three representative  TTS systems: F5-TTS \cite{chen2025f5}, XTTS-v2 \cite{casanova2024xtts}, and CosyVoice2 \cite{du2024cosyvoice}. These models are widely used in downstream fine-tuning settings and collectively cover diverse architectural paradigms, including flow-matching-based, autoregressive, and hybrid designs.
For all experiments, we use publicly available pre-trained checkpoints and follow the official fine-tuning protocols with default hyperparameters. Specifically, F5-TTS uses the F5-TTS-v1-Base model trained on the Emilia dataset \cite{he2024emilia}, XTTS-v2 is pre-trained on a mixture of LibriTTS-R \cite{koizumi2023libritts}, LibriLight \cite{kahn2020libri}, and internal audiobook-style datasets, while CosyVoice2 is initialized from the CosyVoice2-0.5B checkpoint, which is trained on large-scale in-house data.

\noindent\textbf{Fine-tuning Datasets.}
We utilize the VCTK\cite{yamagishi2019cstr} and the British Dialect dataset\cite{demirsahin2020open} for evaluation.

\noindent\textit{VCTK:} This corpus is a multi-speaker dataset designed for the training and evaluation of TTS systems, encompassing over 44,000 distinct audio samples that total approximately 44 hours of recordings collected from 110 English speakers with diverse regional accents. Within this corpus, each speaker recites approximately 400 distinct sentences sourced from newspaper publications, the Rainbow Passage, and a standard elicitation paragraph formulated to ensure comprehensive phonetic coverage. Consequently, this dataset serves as a standard benchmark for the fine-tuning of clean speech.

\noindent\textit{British Dialect Dataset:} This collection encompasses 17,877 transcribed utterances, totaling 31 hours of high-quality audio recordings collected from 120 native speakers. These individuals represent six distinct regional varieties, namely, Southern England, Midlands, Northern England, Welsh, Scottish, and Irish English. The corpus utilizes scripts specifically formulated for accent elicitation to ensure comprehensive phonetic coverage. Furthermore, this serves as an optimal use case for the fine-tuning of models on specific speech variations.

\noindent\textbf{Dataset Split.}
Following the standard partitioning protocol in prior membership inference attacks~\cite{shokri2017membership, duan2023diffusion}, we randomly sample 100 speakers from each dataset and split them into two disjoint 50-speaker sets: $S_{1}$ for fine-tuning the victim model and evaluating the attack, and $S_{2}$ for training the shadow model and the attack classifier. Within each set, speakers are divided into $\textit{spk}_{\text{in}}$ and $\textit{spk}_{\text{out}}$, and the corresponding model (victim or shadow) is fine-tuned on $N$ utterances sampled from $\textit{spk}_{\text{in}}$. For speaker-level MIA, member samples are re-sampled from $\textit{spk}_{\text{in}}$ with a seed different from that of the training set, while non-members are drawn from $\textit{spk}_{\text{out}}$. For record-level MIA, members are the $N$ training utterances, and non-members are another $N$ utterances sampled from the remaining utterances of both speaker groups. We set $N=5000$ for VCTK and $N=3000$ for British Dialect. Unless otherwise stated, speaker-level MIA uses $n = 3$ attacker utterances and each query
is repeated with $m = 5$ random seeds, and $m = 10$ for record-level MIA.

\noindent\textbf{Evaluation Metrics.}
We adopt the standard metrics for evaluating membership inference attacks, including area under the curve (AUC), average-case ''accuracy'' (ACC), and true positive rate versus false positive rate in the low-false positive rate regime (TPR@1\%FPR). Following prior work that argues that meaningful privacy evaluation should emphasize the worstcase~\cite{carlini2022membership}, our primary focus is the TPR-FPR curve in the low false-positive regime, as it better reflects an adversary's ability to confidently identify training members.

\subsection{Attack Performance}
\label{sec:attack_performance}

We evaluate the proposed attack suite on three TTS models and two datasets. The results show that fine-tuned TTS models leak membership information at both levels. We further show that record-level MIA remains effective in a more challenging setting where the model tries to distinguish records of the same speakers, confirming that the two attacks capture distinct forms of memorization.

\subsubsection{TTS Models Are Vulnerable at Both Speaker and Record Levels}
\label{sec:main_results}
\begin{figure*}[t]
  \centering

  \subfloat[Speaker MIA on VCTK.]{%
        \includegraphics[width=0.24\textwidth]{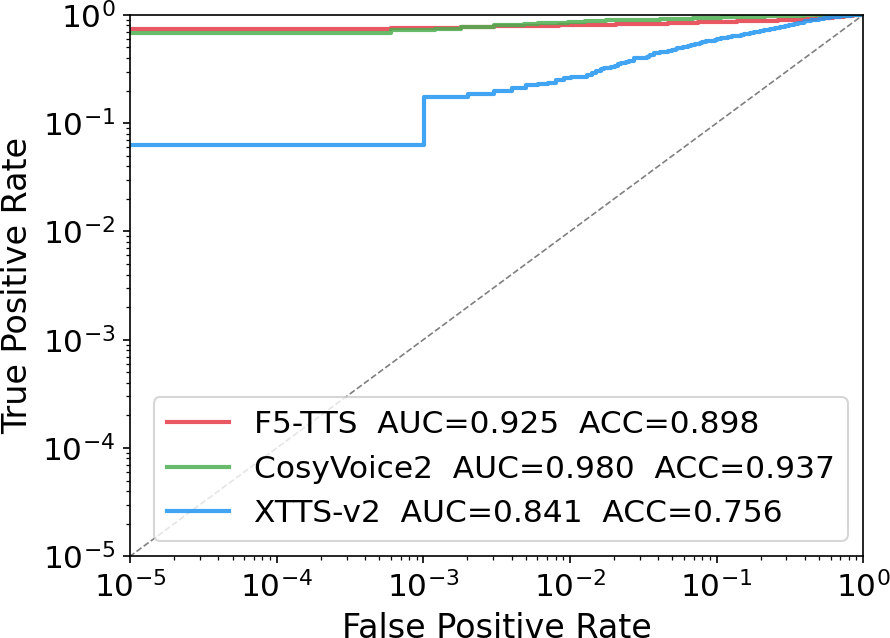}%
        \label{fig:spk_vctk}}
  \hfill
  \subfloat[Speaker MIA on British Dialect.]{%
        \includegraphics[width=0.24\textwidth]{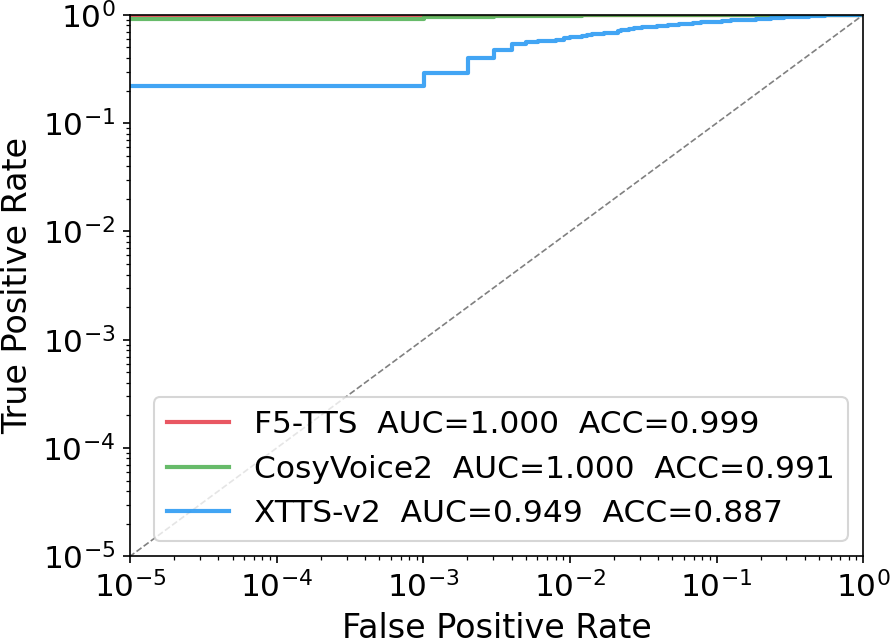}%
        \label{fig:spk_dialect}}
  \hfill
  \subfloat[Record MIA on VCTK.]{%
        \includegraphics[width=0.24\textwidth]{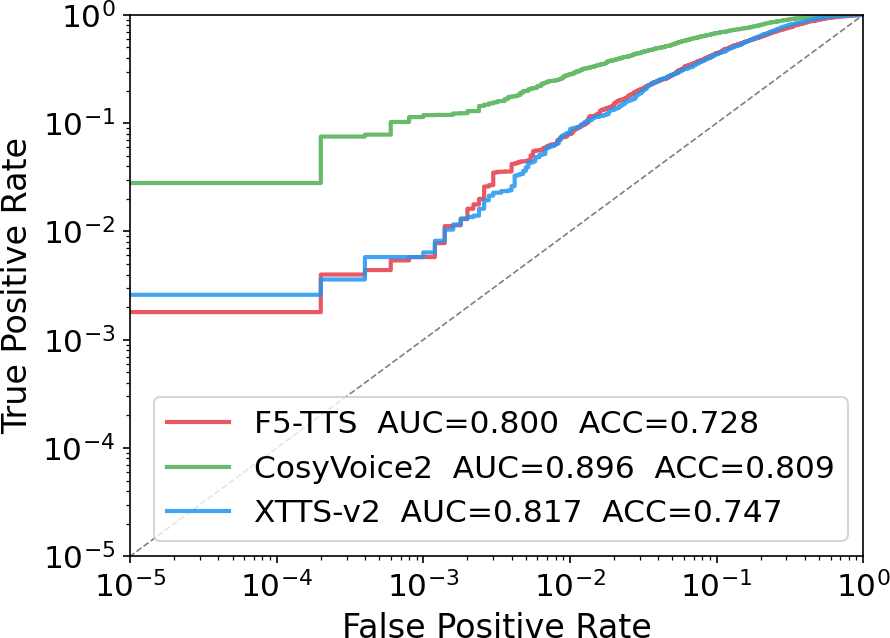}%
        \label{fig:utt_vctk}}
  \hfill
  \subfloat[Record MIA on British Dialect.]{%
        \includegraphics[width=0.24\textwidth]{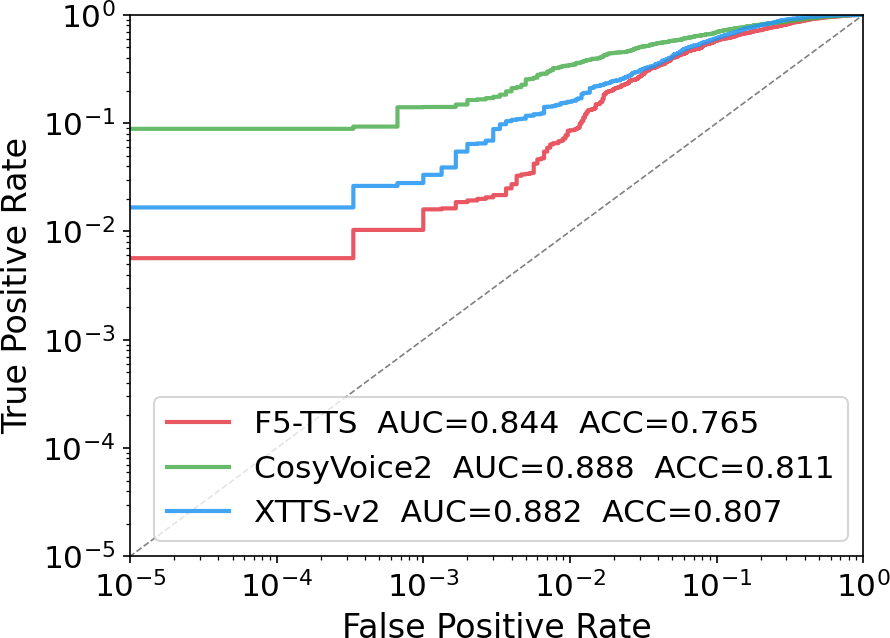}%
        \label{fig:utt_dialect}}

  \caption{Attack performance of speaker-level MIA and record-level MIA on three victim TTS models and two fine-tuning datasets. The diagonal line indicates the random guess baseline.}
  \label{fig:main_results}
\end{figure*}

Figure~\ref{fig:main_results} shows the main attack results. Overall, the attacks are highly effective across all three target models and both datasets. Figure~\ref{fig:spk_vctk} and Figure~\ref{fig:spk_dialect} show that Speaker-level MIA is consistently strong, with all AUC values above 0.8 and near-perfect AUC in the strongest settings, namely CosyVoice2 and F5-TTS on the British Dialect dataset. The same trend is visible in the low-FPR region, where TPR remains substantially above chance even under strict false-positive constraints. As is shown in Figure~\ref{fig:utt_vctk} and Figure~\ref{fig:utt_dialect}, Record-level MIA is also effective across all models and datasets, with AUC values consistently well above the random-guess baseline of 0.5, although it is weaker than speaker-level MIA. This gap likely arises because speaker membership is reflected in stable identity patterns that persist across utterances, whereas record-level attacks must detect finer content-dependent memorization traces that are more easily weakened by variation in the generation process.

The attack performance also differs across datasets. In general, British Dialect yields stronger attack results than VCTK at both the speaker and record levels. This difference is especially clear for speaker-level MIA, where all three target models achieve higher AUC and TPR@1\%FPR on British Dialect (See Figures~\ref{fig:spk_vctk} and~\ref{fig:spk_dialect}). At the record level, the gap is smaller, but CosyVoice2 and XTTS-v2 still show stronger low-FPR performance on British Dialect (See Figures~\ref{fig:utt_vctk} and~\ref{fig:utt_dialect}). One possible reason is the richer accent and speaking-style variation in British Dialect: more acoustically distinctive speakers may leave memorization signals that remain more separable after generation.

Attack effectiveness also varies across model architectures, and the vulnerability ranking differs between the two attack levels. For speaker-level MIA, XTTS-v2 is consistently the least vulnerable, with AUC of 0.841 on VCTK and 0.949 on British Dialect (See Figures~\ref{fig:spk_vctk} and~\ref{fig:spk_dialect}). We attribute this to the fact that XTTS-v2 fine-tuning does not directly adapt the acoustic decoder that maps latent representations to mel-spectrograms or raw waveforms, leaving fewer speaker-identity traces in the generated output. For record-level MIA, F5-TTS yields the lowest AUC on both datasets, achieving 0.800 on VCTK and 0.844 on British Dialect (See Figures~\ref{fig:utt_vctk} and~\ref{fig:utt_dialect}). A likely reason is that F5-TTS lacks autoregressive modeling, which means that it does not capture the sequential and local content dependencies required by record-level MIA to distinguish memorized records from non-members. Our ablation on the hybrid architecture of CosyVoice2 (Section~\ref{sec:model_component_comparison}) further supports this, showing that speaker- and record-level leakage are driven by different model components.

\begin{figure}[!htbp]
  \centering
  \includegraphics[width=0.6\linewidth]{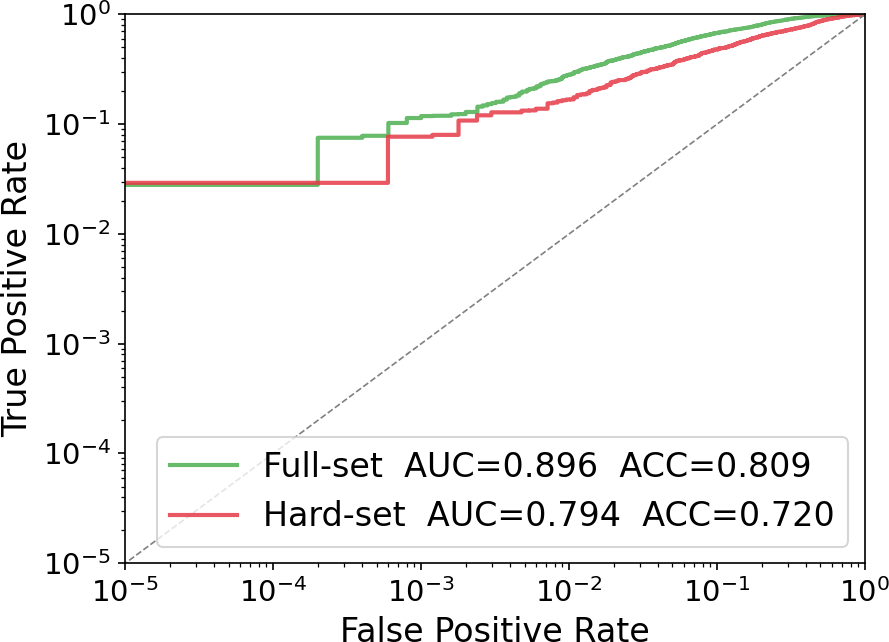}
  \caption{Record-level MIA maintains comparable low-FPR performance even when non-members are drawn from seen speakers, confirming that the attack captures record-specific memorization beyond speaker identity.}
  \label{fig:utt_hard}
\end{figure}

\subsubsection{Record-Level MIA Captures Record-Specific Memorization Beyond Speaker Identity.}
We further evaluate a more challenging setting in which all non-member records are drawn from speakers who appear in the fine-tuning set. In this setting, the attack must distinguish between different records from the same speakers, rather than simply separating seen speakers from unseen ones. As shown in Figure~\ref{fig:utt_hard}, the ROC curve remains clearly above the random baseline, with low-FPR performance comparable to that on the full evaluation set. This result suggests that record-level MIA captures record-specific memorization cues beyond speaker identity.

\subsection{Ablation and Analysis}
\label{sec:ablation_study}

In this section, we provide the ablation results on the key design choices of the proposed attack suite, examining the effect of query strategy, feature representation, temporal alignment, fine-tuned model components, and query budget on attack performance.

\subsubsection{Recitation Is the Strongest Overall Query}
\label{sec:ablation_query}
\begin{figure}[t]
  \centering
  \subfloat[Speaker-Level MIA.]{%
        \includegraphics[width=0.49\linewidth]{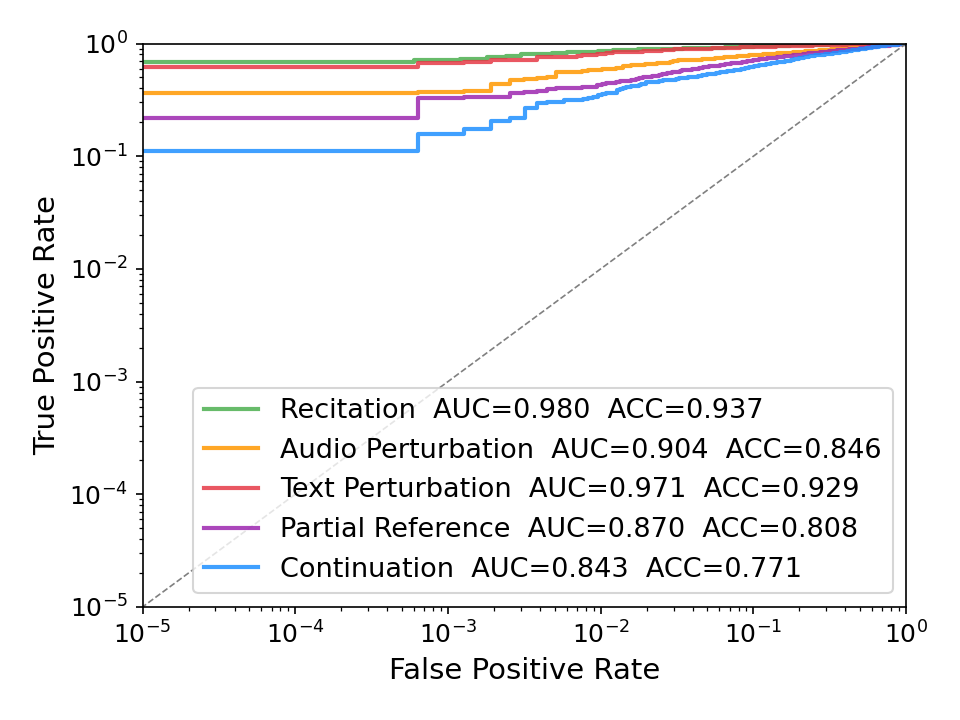}%
        \label{fig:query_cosyvoice}}
  \hfill
  \subfloat[Record-Level MIA.]{%
        \includegraphics[width=0.49\linewidth]{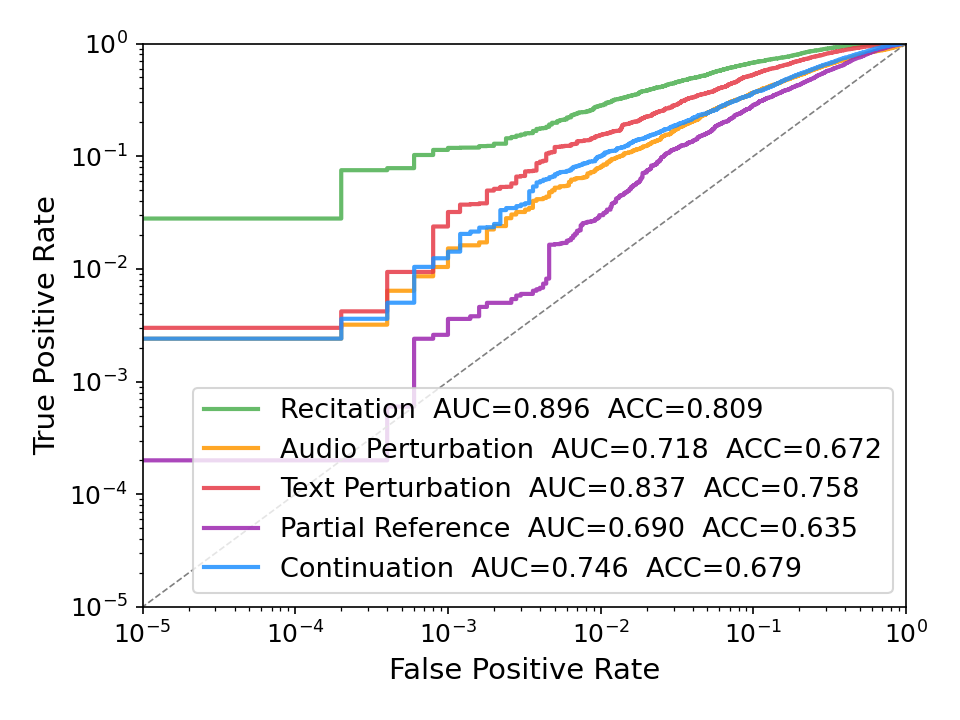}%
        \label{fig:utt_query_cosyvoice}}
  \caption{Recitation best exposes membership information at both levels, with a larger margin for record-level MIA.}
  \label{fig:text_query_comparison}
  \label{fig:utt_query_comparison}
\end{figure}

Figure~\ref{fig:text_query_comparison} compares the five queries in Table~\ref{tab:query_space} using CosyVoice2 fine-tuned on VCTK. The concrete configuration of each query is given in Appendix~\ref{app:query_config}. Recitation, which ranks first under both criteria (Section~\ref{sec:design_principle}), performs best at both levels, reaching an AUC of 0.980 for speaker-level MIA and 0.896 for record-level MIA. Its margin over the second-best query is small at the speaker level (0.009 AUC) but substantial at the record level (0.059 AUC), and larger still in the low-FPR regime, where its TPR@1\%FPR nearly doubles that of the second-best query (0.281 vs.\ 0.154).

The two levels differ in which queries approach recitation, in line with the predictions in Section~\ref{sec:design_principle}. For speaker-level MIA, text perturbation performs nearly as well as recitation (0.971 vs.\ 0.980 AUC; 0.802 vs.\ 0.853 TPR@1\%FPR). It supplies the complete, clean $s^{*}$ and thus retains the full elicitation cue under C2, and the replaced words matter little at this level, where identity extraction depends only weakly on the exact linguistic content. For record-level MIA, performance instead follows how completely a query satisfies C2. The two queries that supply the complete, clean reference (recitation and text perturbation) perform best (0.896 and 0.837 AUC; 0.281 and 0.154 TPR@1\%FPR), whereas the three that supply an incomplete or degraded reference fall to an AUC of 0.690--0.746 and a TPR@1\%FPR of 0.030--0.100. Text perturbation nonetheless remains below recitation at this level, as the replaced words reduce the scorable extent under C1. The record-level comparison is thus far more sensitive than the speaker-level to the completeness of the reference speech.

\subsubsection{Speaker- and Record-Level Attacks Require Features at Different Granularities}

\label{sec:feature_comparison}
\begin{figure}[t]
  \centering
  \subfloat[Speaker-Level MIA.]{%
        \includegraphics[width=0.49\linewidth]{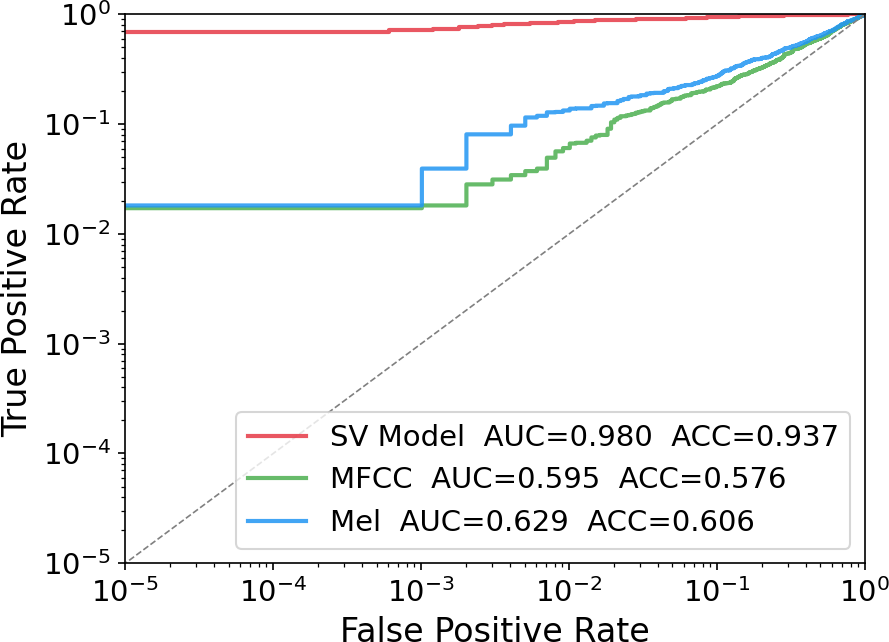}%
        \label{fig:feats_spk}}
  \hfill
  \subfloat[Record-Level MIA.]{%
        \includegraphics[width=0.49\linewidth]{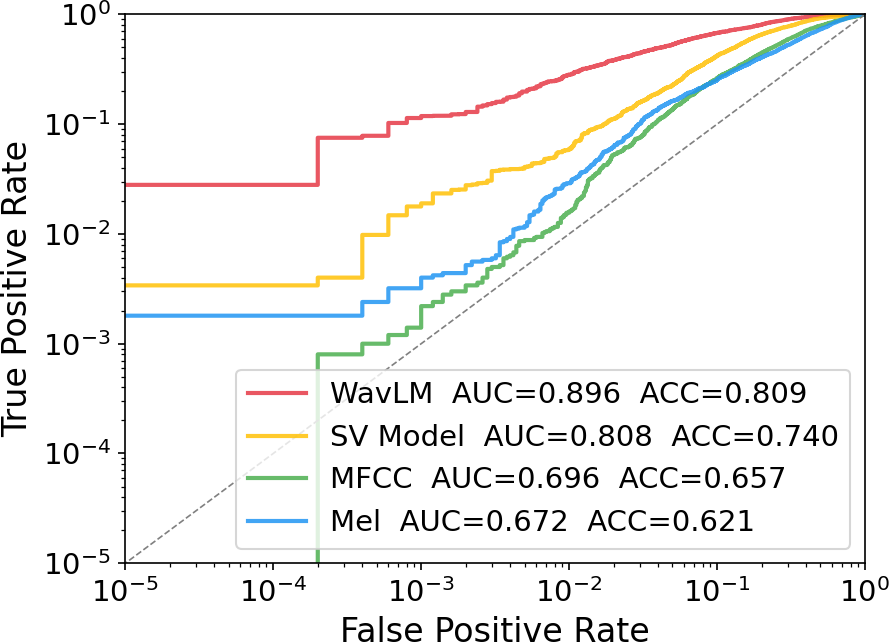}%
        \label{fig:feats_utt}}
  \caption{SV features best expose speaker-level leakage, while WavLM best exposes record-level leakage.}
  \label{fig:mia_feature_comparison}
\end{figure}

Figure~\ref{fig:mia_feature_comparison} compares the features used to compute attack scores for speaker-level and record-level MIA. For speaker-level MIA, as shown in Figure~\ref{fig:feats_spk}, the SV representation is the most effective, achieving an AUC of 0.980. It substantially outperforms Mel and MFCC, which achieve AUC values of 0.629 and 0.595, respectively, with an even larger gap in the low-FPR region. This improvement is because the SV model is explicitly trained to suppress linguistic and prosodic variation while producing global, utterance-invariant embeddings that preserve speaker identity.

In contrast, record-level MIA requires more fine-grained features. As shown in Figure~\ref{fig:feats_utt}, the SV representation alone is insufficient, achieving an AUC of 0.808, whereas multi-level WavLM features aggregated by the attack classifier improve the AUC to 0.896. Both representations substantially outperform Mel and MFCC, which achieve AUC values of 0.672 and 0.696, respectively. This gap indicates that record-level membership depends on local, content-dependent acoustic patterns that compact speaker embeddings discard by design. The contrast between the two panels further shows that speaker-level and record-level leakage arise at different granularities: global identity cues capture speaker-level leakage, whereas local sequence-specific cues capture record-level leakage.

\subsubsection{Temporal Concatenation Captures More Stable Speaker Identity.}
\label{sec:concat_vs_avg}
\begin{figure}[t]
  \centering
  \subfloat[XTTS-v2.]{%
        \includegraphics[width=0.49\linewidth]{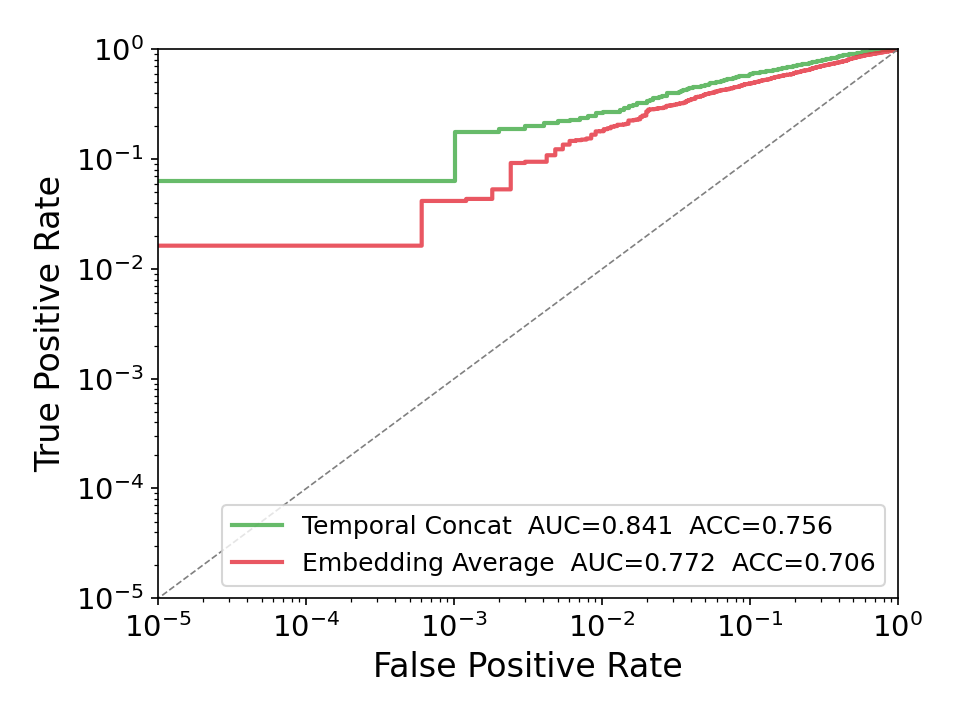}%
        \label{fig:concat_vs_avg_xtts}}
  \hfill
  \subfloat[CosyVoice2.]{%
        \includegraphics[width=0.49\linewidth]{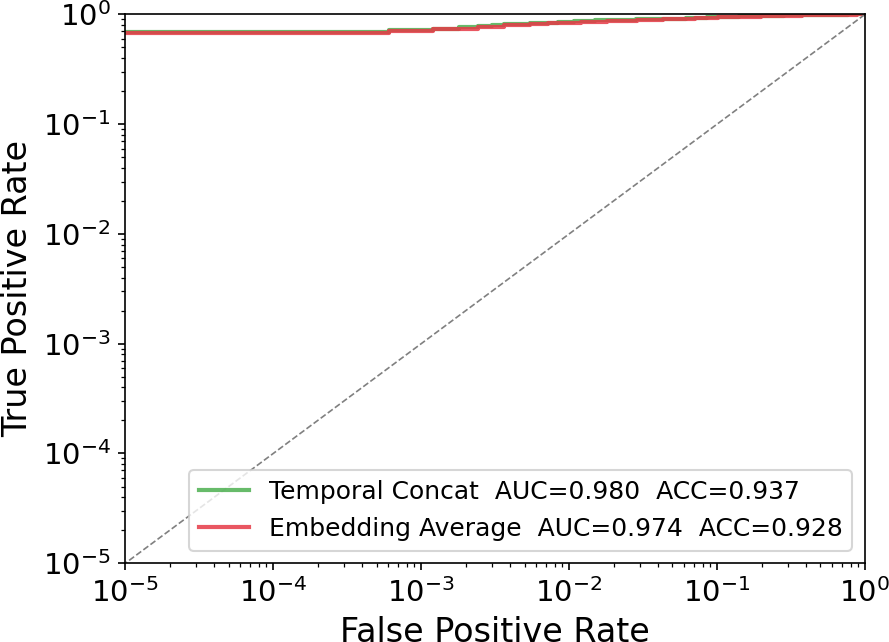}%
        \label{fig:concat_vs_avg_cosyvoice}}
  \caption{Temporal concatenation consistently outperforms embedding averaging for speaker-level MIA, with the largest gains on the more challenging XTTS-v2 model.}
  \label{fig:concat_vs_avg_emb}
\end{figure}

Figure~\ref{fig:concat_vs_avg_emb} shows that temporal concatenation yields more stable speaker representations than per-clip embedding averaging for speaker-level MIA. As shown in Figure~\ref{fig:concat_vs_avg_xtts}, on XTTS-v2, concatenation achieves an AUC of 0.841 and an ACC of 0.756, compared with 0.772 and 0.706 for the averaging baseline, respectively. The gap is especially prominent at low FPR, where concatenation maintains a higher TPR across the low false-positive regime. As shown in Figure~\ref{fig:concat_vs_avg_cosyvoice}, on CosyVoice2, the margin narrows because both strategies already perform well, but concatenation still achieves a higher AUC of 0.980, compared with 0.974 for embedding averaging.

\subsubsection{Record-Level MIA Requires Multi-Level WavLM Features.}
\label{sec:layer_analysis}
\begin{figure}[t]
  \centering
  \subfloat[MIA performance with each individual WavLM layer feature; lower layers are generally more discriminative than upper layers.]{%
        \includegraphics[width=\linewidth]{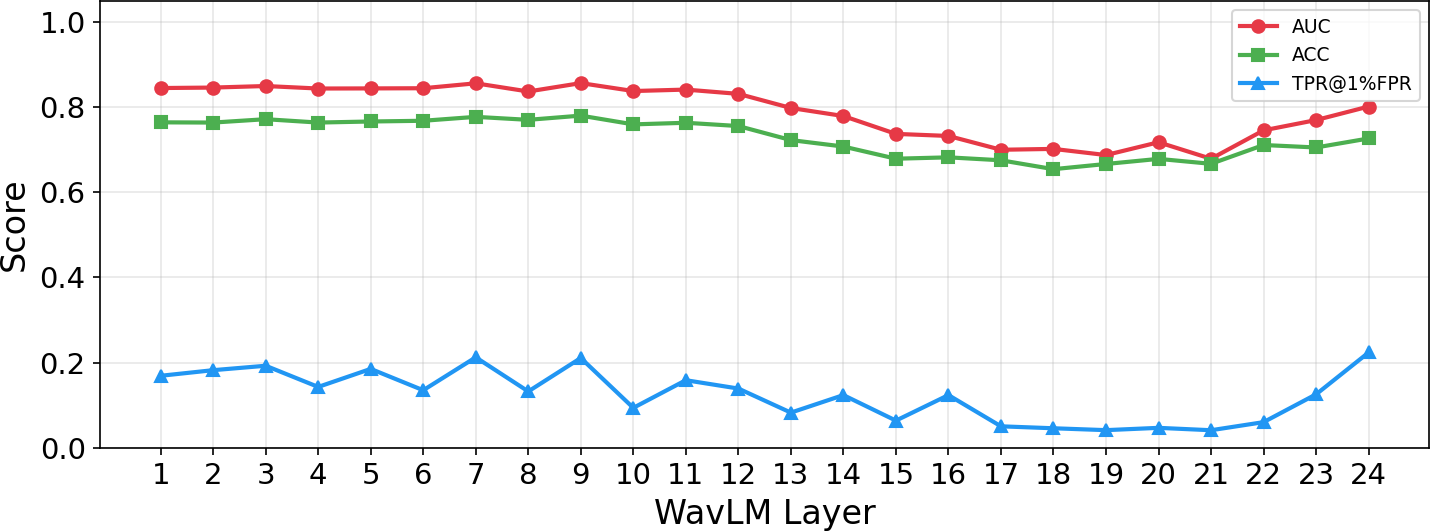}%
        \label{fig:utt_wavlm_single_layer}}

  \subfloat[Aggregating more top-ranked layers improves MIA performance monotonically, making full-hierarchy aggregation the best default.]{%
        \includegraphics[width=\linewidth]{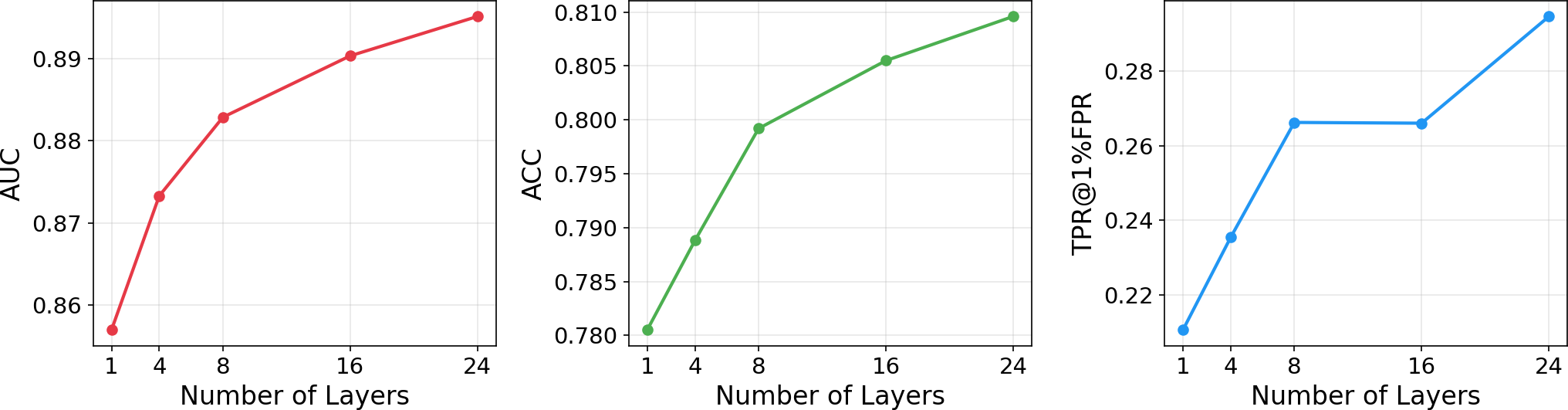}%
        \label{fig:utt_wavlm_topk_layer}}
  \caption{Multi-level features improve record-level MIA.}
  \label{fig:utt_wavlm_layer_combined}
\end{figure}

To further examine the role of multi-level information in record-level MIA, Figure~\ref{fig:utt_wavlm_layer_combined} analyzes which WavLM layers are most useful and whether aggregating layers improves over using a single layer. The best single-layer results come from the lower WavLM hierarchy, where the first twelve layers consistently outperform most higher layers in AUC and ACC. TPR@1\%FPR is noisier, but it also tends to peak in the lower layers. This suggests that record-level memorization relies more on fine-grained local acoustic matching than on the more abstract semantic representations encoded in upper layers. The signal, however, is not limited to one layer. Combining top-ranked layers steadily improves all reported metrics, with AUC increasing from roughly 0.857 for the best single layer to 0.873, 0.883, 0.890, and 0.896 when aggregating the top-4, top-8, top-16, and all 24 layers, respectively. ACC and TPR@1\%FPR follow the same trend. These results suggest that different layers provide complementary information, and pooling them with the LSTM aggregator improves membership discrimination. This makes full-hierarchy aggregation a strong default choice for record-level MIA.

\subsubsection{Record-Level MIA Requires Fine-Grained Temporal Alignment}
\label{sec:alignment}
\begin{figure}[t]
  \centering
  \includegraphics[width=0.60\linewidth]{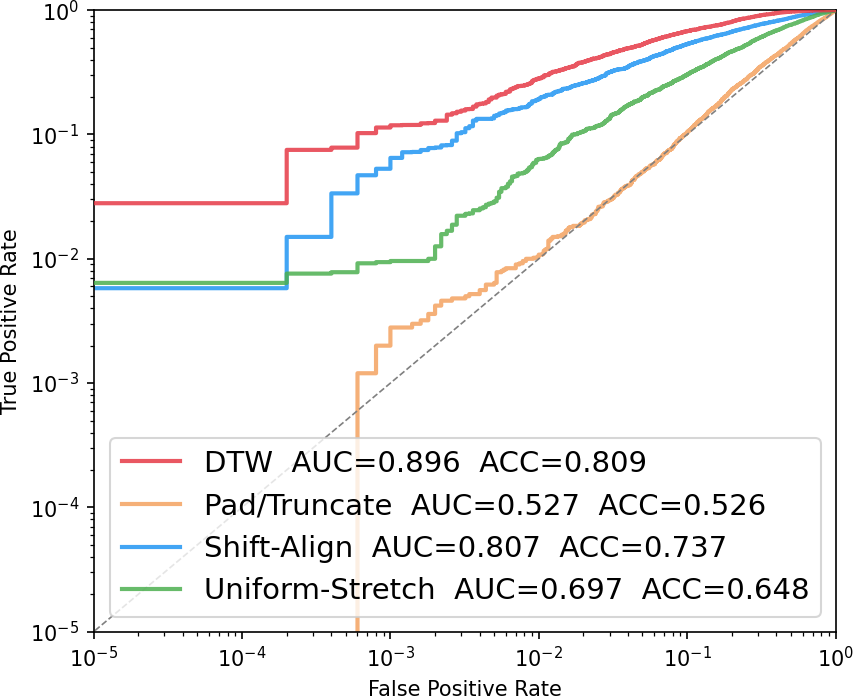}
  \caption{DTW substantially outperforms alternative alignment methods for record-level MIA, showing that fine-grained local alignment is important.}
  \label{fig:alignment_method}
\end{figure}

Figure~\ref{fig:alignment_method} empirically validates the alignment design introduced in Section~\ref{sec:representation_learning}. A naive baseline that pads or truncates the generated sequence to match the reference length, without performing explicit alignment, yields near-random performance. This result confirms that temporal alignment is necessary for meaningful comparison. DTW achieves the strongest performance, with an AUC of 0.896 and an ACC of 0.809. It clearly outperforms shift-align, which corrects only a fixed temporal offset between the reference and generated sequences and reaches an AUC of 0.807 and an ACC of 0.737, as well as uniform-stretch, which globally rescales the generated sequence to match the reference duration and achieves an AUC of 0.697 and an ACC of 0.648. The large gap between DTW and these simpler alternatives indicates that record-level MIA cannot be handled by correcting only a global shift or a uniform temporal rescaling. Instead, the attack requires fine-grained alignment between phonetically corresponding regions so that local acoustic matches are preserved rather than treated as temporal mismatch.

\subsubsection{Speaker and Record-Level Leakage Are Driven by Different Model Components}
\label{sec:model_component_comparison}

\begin{figure}[t]
  \centering
  \subfloat[Speaker-Level MIA.]{%
        \includegraphics[width=0.48\linewidth]{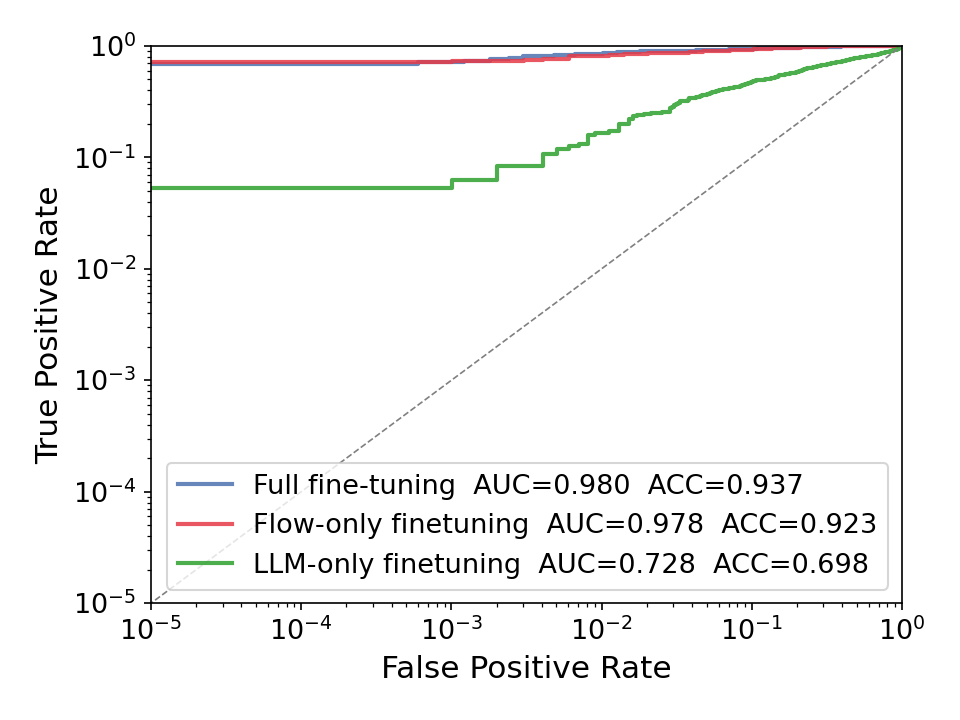}%
        \label{fig:spk_model_comp}}
  \hfill
  \subfloat[Record-Level MIA.]{%
        \includegraphics[width=0.48\linewidth]{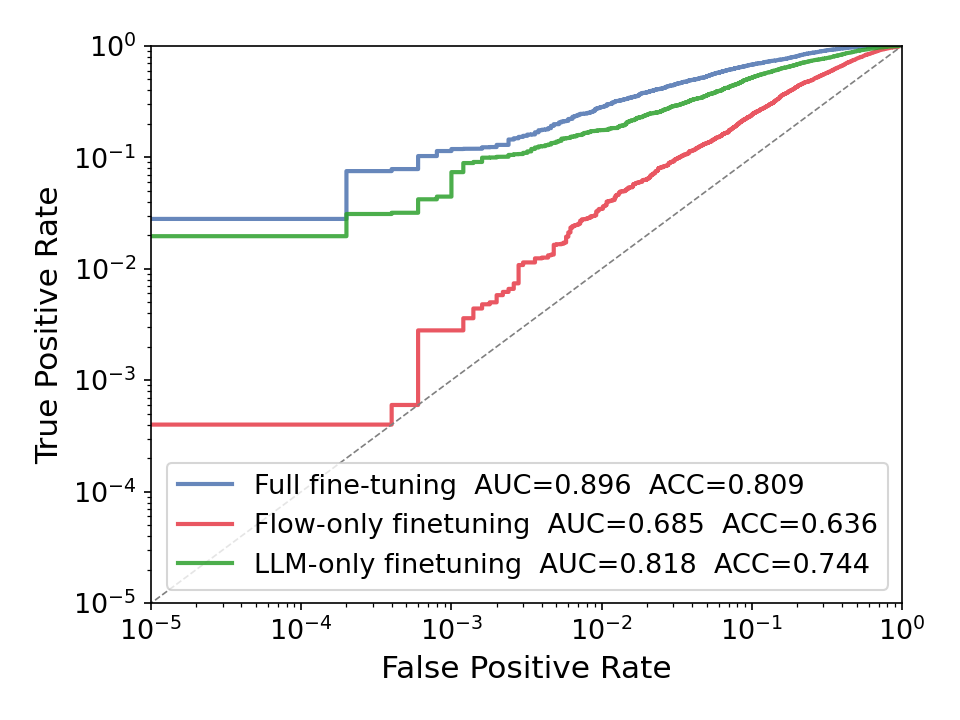}%
        \label{fig:utt_model_comp}}
  \caption{Speaker-level leakage is primarily driven by the flow-matching acoustic decoder, whereas record-level leakage relies more on the LLM component, suggesting different underlying memorization mechanisms.}
  \label{fig:model_component_comparison}
\end{figure}

Since CosyVoice2 consists of an autoregressive language model over latent audio tokens and a flow-matching acoustic decoder that maps these tokens to mel-spectrograms, we examine the effect of fine-tuning each component on speaker-level and record-level MIA, as shown in Figure~\ref{fig:model_component_comparison}. For speaker-level MIA, Figure~\ref{fig:spk_model_comp} shows that full and flow-only fine-tuning produce near-perfect AUC and high TPR at low FPRs. In contrast, LLM-only fine-tuning drops substantially to an AUC of 0.728, with a corresponding collapse in low-FPR TPR. The minimal gap between full and flow-only fine-tuning, together with the large drop under LLM-only fine-tuning, suggests that speaker-level leakage is mainly driven by the acoustic decoder.

For record-level MIA in Figure~\ref{fig:utt_model_comp}, full fine-tuning again achieves the strongest performance, with an AUC of 0.896, but the pattern under partial fine-tuning differs from the speaker-level case. LLM-only fine-tuning retains substantial leakage, with an AUC of 0.818, whereas flow-only fine-tuning drops further to an AUC of 0.685. The same ordering is also reflected in TPR in the low-FPR region. This trend is the reverse of the speaker-level result, where flow-only fine-tuning nearly matches full fine-tuning. One possible interpretation is that record-level memorization relies more on the autoregressive component, which captures prosodic and temporal traces, whereas the acoustic decoder primarily encodes identity.

\subsubsection{Speaker-Level MIA Improves with More Attacker Records and Converges at 7--9}
\label{sec:spk_nutt_ablation}

\begin{figure}[t]
  \centering
  \includegraphics[width=0.60\linewidth]{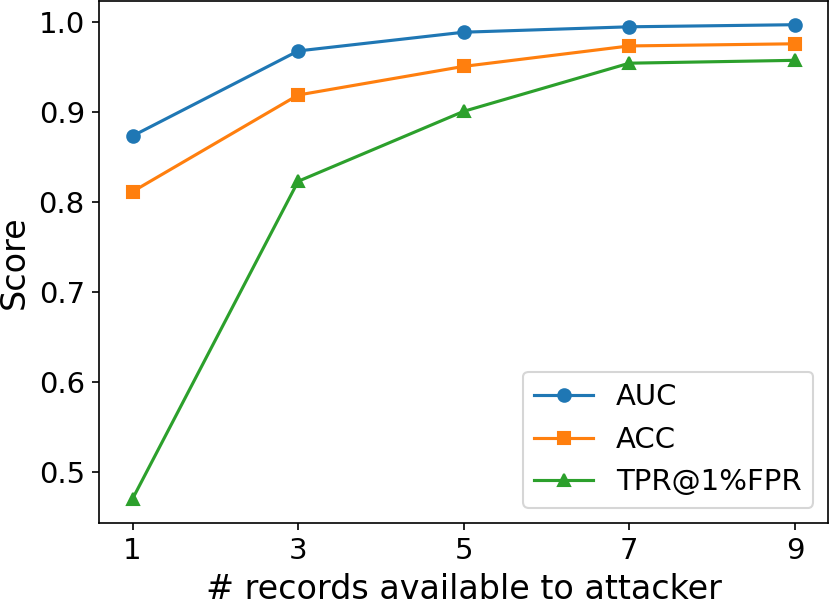}
  \caption{Speaker-Level MIA performance improves with the number of attacker utterances and
converges at around 7--9 recordings.}
  \label{fig:spk_nutt_ablation}
\end{figure}

Figure~\ref{fig:spk_nutt_ablation} shows that speaker-level MIA improves steadily as more
recordings of the target speaker become available to the attacker. Increasing the number of
records from 1 to 3 raises the AUC from approximately 0.87 to 0.96 and the
TPR@1\%FPR from about 0.47 to 0.82. Performance continues to improve with additional
utterances and converges at around 7--9 records, where the AUC approaches 1.0 and the
TPR@1\%FPR stabilizes near 0.95. This trend follows from the fact that concatenating more
utterances yields a longer and phonetically richer input to the SV encoder, producing a more
stable speaker embedding. In practice, an auditor with access to fewer than ten public
recordings of a target speaker can reliably determine whether the voice of that speaker was
used for fine-tuning.

\subsubsection{Effect of Query Budget}
\label{sec:query_budget}
\begin{figure}[t]
  \centering
  \subfloat[Speaker-Level MIA.]{%
        \includegraphics[width=0.49\linewidth]{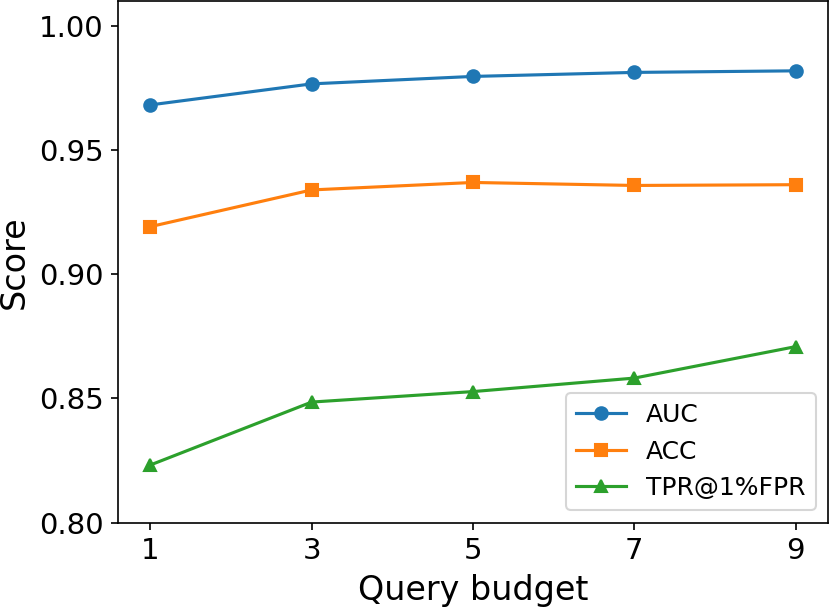}%
        \label{fig:spk_query_budget}}
  \hfill
  \subfloat[Record-Level MIA.]{%
        \includegraphics[width=0.49\linewidth]{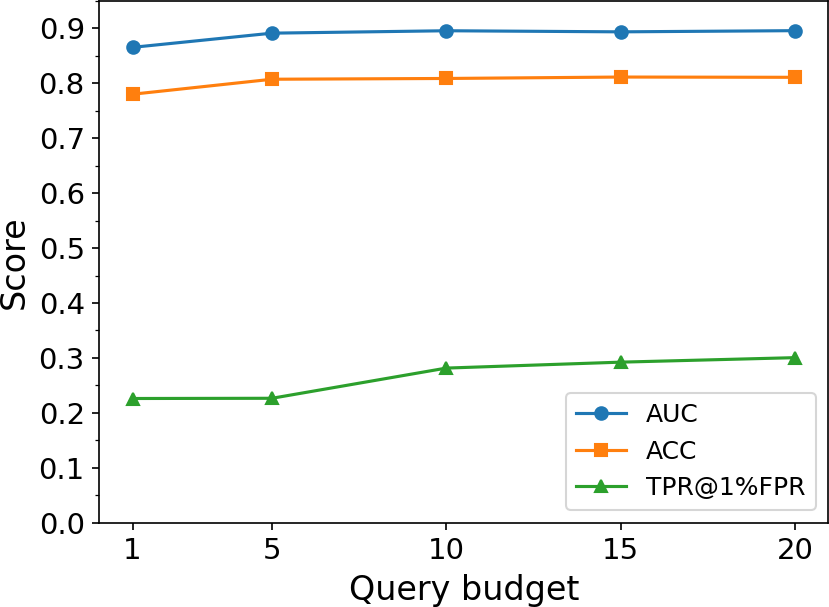}%
        \label{fig:utt_query_budget}}
  \caption{Effect of query budget, defined as the number of repeated queries with different random seeds, on speaker-level and record-level MIA.}
  \label{fig:query_budget}
\end{figure}

Figure~\ref{fig:spk_query_budget} shows that the performance of speaker-level MIA improves as the number of repeated queries increases, but most of the gain appears early. The AUC rises from 0.968 with one seed to 0.980 with five seeds and then plateaus, while TPR@1\%FPR increases more gradually from 0.823 to 0.871 by nine seeds. The early saturation in AUC indicates that average-case classification becomes reliable with few repetitions; additional repetitions mainly refine the low-FPR operating point, where score variance matters most.

Figure~\ref{fig:utt_query_budget} shows a similar but slower saturation pattern for record-level MIA. The AUC improves from roughly 0.863 with one seed to 0.896 with ten or more seeds, and TPR@1\%FPR increases from about 0.23 to 0.30 by twenty seeds. Compared with speaker-level MIA, record-level scores are noisier for each query because content-specific memorization traces are weaker. More repeated queries are therefore needed before the low-FPR regime stabilizes.

\subsection{Data Characteristics and Vulnerability}
\label{sec:speech_factors}

Beyond aggregate attack performance, we examine which speech properties affect vulnerability to membership inference, revealing what information the model memorizes from the training data. For speaker-level MIA, we find that male speakers are more vulnerable than female speakers. At the record level, utterances with lower silence ratio, higher phoneme density, greater high-frequency energy and longer duration are more vulnerable. Unless otherwise noted, all analyses use CosyVoice2 on VCTK.

\subsubsection{Male Speakers Are More Vulnerable to Speaker-Level MIA}
\begin{figure}[t]
  \centering
  \includegraphics[width=\linewidth]{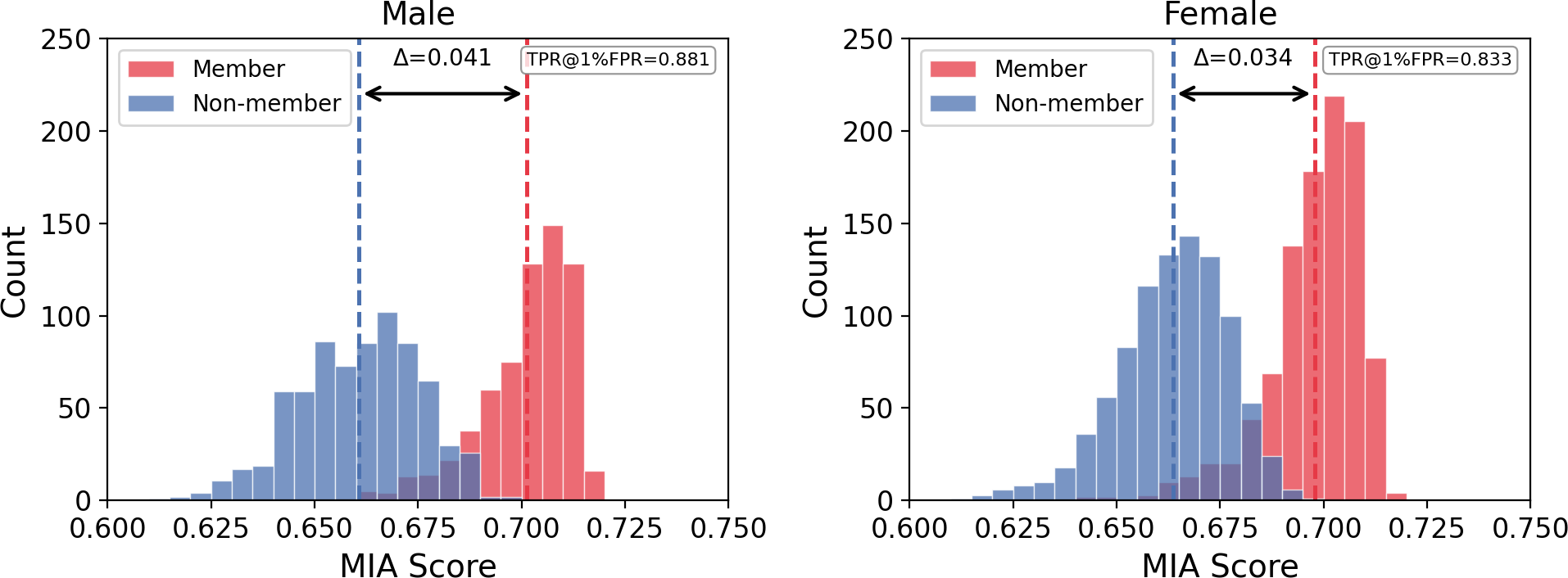}
  \caption{Male speakers are more vulnerable to speaker-level MIA than female speakers, possibly due to greater within-group acoustic variability in female speech.}
  \label{fig:gender_vulnerability_spk_mia}
\end{figure}

Figure~\ref{fig:gender_vulnerability_spk_mia} shows that male speakers are more vulnerable to Speaker-Level MIA than female speakers. Specifically, the separation between the member and non-member score distributions is larger for male speakers, with a mean score gap of $\Delta=0.041$ compared with $\Delta=0.034$ for female speakers. This larger gap is also reflected in the low-FPR operating regime, where the TPR@1\%FPR reaches 0.881 for male speakers but 0.833 for female speakers. A possible explanation is that female speech exhibits larger within-group acoustic variability, which leads to greater overlap between member and non-member score distributions. 
Prior work has reported higher within-group variation for female voices and has noted that female speakers are often more difficult to discriminate in speaker recognition~\cite{childers1991gender}. This may reduce the separability of the membership signal for female speakers and partially explain why male speakers appear more vulnerable to Speaker-Level MIA (easier to distinguish members from non-members) in our results.

\subsubsection{Lower Silence Ratio Increases Record-Level Vulnerability}
\begin{table}[t]
\small
\centering
\caption{Record-level MIA performance across silence ratio bins; Lower Silence Ratio Increases Record's Risk.}
\label{tab:silence_ratio_bins}
\setlength{\tabcolsep}{8pt}
\begin{tabular}{lccc}
\toprule
\textbf{Bin} & \textbf{AUC} & \textbf{ACC} & \textbf{TPR@1\%FPR} \\
\midrule
0--20\%   & 0.943 & 0.886 & 0.406 \\
20--30\%  & 0.936 & 0.859 & 0.427 \\
30--40\%  & 0.919 & 0.828 & 0.344 \\
40--50\%  & 0.886 & 0.793 & 0.258 \\
50--60\%  & 0.850 & 0.741 & 0.246 \\
$>60\%$   & 0.748 & 0.625 & 0.129 \\
\bottomrule
\end{tabular}
\end{table}

Silence ratio measures the fraction of an utterance occupied by non-speech or low-energy regions. Table~\ref{tab:silence_ratio_bins} shows a clear monotonic decline in attack performance as the silence ratio increases. The AUC drops from 0.943 in the 0--20\% bin to 0.748 in the $>60\%$ bin, while TPR@1\%FPR decreases from 0.406 to 0.129. We attribute this trend to the fact that silence provides little discriminative acoustic information for comparison. As the silence ratio increases, the amount of informative content available to distinguish members from non-members correspondingly decreases.

\subsubsection{Denser Phoneme Structure Increases Record-Level Vulnerability}
\begin{table}[t]
\small
\centering
\caption{Record-level MIA performance across phoneme density bins; higher phoneme density can increases risk.}
\label{tab:phoneme_density_bins}
\setlength{\tabcolsep}{8pt}
\begin{tabular}{lccc}
\toprule
\textbf{Bin} & \textbf{AUC} & \textbf{ACC} & \textbf{TPR@1\%FPR} \\
\midrule
$<$3                 & 0.852 & 0.759 & 0.184 \\
3--3.5             & 0.888 & 0.794 & 0.262 \\
3.5--4             & 0.907 & 0.807 & 0.335 \\
4--4.5             & 0.927 & 0.828 & 0.372 \\
$>4.5$             & 0.912 & 0.819 & 0.340 \\
\bottomrule
\end{tabular}
\end{table}

Phoneme density measures how many phonemes on average appear per syllable in an utterance and serves as a proxy for articulatory complexity. Table~\ref{tab:phoneme_density_bins} shows that utterances with denser phoneme structure are generally more vulnerable to MIA. AUC rises from 0.852 in the $<3$ bin to a peak of 0.927 in the 4--4.5 bin, and TPR@1\%FPR increases from 0.184 to 0.372 over the same range. Phonemically denser utterances contain richer acoustic transitions and more fine-grained local structure, which provide stronger signals for record-level comparison. The highest-density bin ($>4.5$) shows a slight decline to AUC 0.912 and TPR@1\%FPR 0.340. One possible explanation for this drop is that extremely high phoneme density is associated with fast or heavily clustered articulation, where the synthesis model may smooth over fine-grained detail regardless of membership status, reducing the separability of member and non-member outputs.

\subsubsection{Richer High-Frequency Content Increases Record-Level Vulnerability}
\begin{table}[t]
\small
\centering
\caption{Record-level MIA performance across high-frequency energy ratio bins; richer high-frequency content consistently increases vulnerability.}
\label{tab:hf_ratio_bins}
\setlength{\tabcolsep}{8pt}
\begin{tabular}{lccc}
\toprule
\textbf{Bin} & \textbf{AUC} & \textbf{ACC} & \textbf{TPR@1\%FPR} \\
\midrule
0--1\%   & 0.879 & 0.775 & 0.248 \\
1--3\%   & 0.892 & 0.795 & 0.295 \\
3--6\%   & 0.894 & 0.800 & 0.298 \\
6--12\%  & 0.911 & 0.824 & 0.293 \\
$>12\%$  & 0.937 & 0.853 & 0.438 \\
\bottomrule
\end{tabular}
\end{table}

High-frequency energy ratio measures the proportion of spectral energy above 4\,kHz, which often captures fricatives, sibilants, and other fine acoustic details. Prior work suggests that TTS models have greater difficulty reproducing such high-frequency detail~\cite{bak2023avocodo,ju2024naturalspeech}, which can make these regions more informative for distinguishing member and non-member utterances. Table~\ref{tab:hf_ratio_bins} shows that utterances with higher high-frequency energy ratios are consistently more vulnerable to membership inference. The AUC rises from 0.879 in the 0--1\% bin to 0.937 in the $>12\%$ bin, and TPR@1\%FPR increases from 0.248 to 0.438. One possible explanation is that faithful reproduction of these details may require stronger memorization of the specific recording, which increases the separation between member and non-member outputs.

\subsubsection{Longer Recordings Are More Vulnerable}
\begin{table}[t]
\small
\centering
\caption{Record-level MIA performance across utterance-duration bins; longer utterances show higher vulnerability.}
\label{tab:duration_bins}
\setlength{\tabcolsep}{8pt}
\begin{tabular}{lccc}
\toprule
\textbf{Bin} & \textbf{AUC} & \textbf{ACC} & \textbf{TPR@1\%FPR} \\
\midrule
$<2$s   & 0.857 & 0.736 & 0.282 \\
2--3s   & 0.881 & 0.788 & 0.271 \\
3--4s   & 0.902 & 0.813 & 0.261 \\
4--5s   & 0.912 & 0.819 & 0.320 \\
$>5$s   & 0.916 & 0.823 & 0.395 \\
\bottomrule
\end{tabular}
\end{table}

Utterance duration is the total length of the speech sample in seconds, including both voiced regions and pauses. Table~\ref{tab:duration_bins} shows a monotonically increasing relationship between utterance duration and attack performance. AUC rises steadily from 0.857 for utterances under 2 seconds to 0.916 for those over 5 seconds, and TPR@1\%FPR increases from 0.282 to 0.395 over the same range. A likely explanation is that longer utterances provide more acoustic content for the DTW-based scoring pipeline to work with, yielding more reliable frame-level alignment between the target and generated audio and thus a stronger membership signal.

\section{Discussion}

\noindent\textbf{Broader Impact on Multimodal Generative Models.}
The success of our attack stems from three key principles: matched queries that preserve information and maximize the membership signal, multi-level representations that capture identity and content traces, and temporal alignment for fine-grained comparison of variable-length outputs. These principles also apply to a broader class of generative models with multimodal inputs and continuous, variable-length outputs, including video editing systems~\cite{kuanyv2v} and omni-modal models~\cite{xu2025qwen3,xie2024mini} that jointly process speech, vision, and text.

\noindent\textbf{Potential Defenses and Their Limitations.} 
To mitigate our proposed MIA, defenders might consider empirical mitigations at both the training and inference stages. During training, we explored early stopping to prevent the model from overfitting to the fine-tuning data. While this approach reduces the attack success rate to some extent, it is far from a perfect defense, as the model still memorizes sufficient identity and content traces before convergence. At the inference (system) level, an API provider might accidentally or deliberately perturb the input reference audio (e.g., adding background noise or truncating the audio length) to disrupt our precise temporal alignment. However, our additional experiments show that such naive input perturbations also fail to thwart the attack: even when the input audio is corrupted with noise or halved in length, our MIA still successfully identifies members. Detailed results for these defenses are provided in Sections~\ref{sec:defense_early_stop} and~\ref{sec:defense_input_perturb}.

The insufficiency of these empirical mitigations highlights the robustness of our attack and motivates the necessity of rigorous, albeit costly, provable defenses. To this end, we evaluate DP-SGD~\cite{abadi2016deep} for privacy-preserving TTS fine-tuning on CosyVoice2 with VCTK. Our results demonstrate that DP-SGD with privacy budgets $\varepsilon=4$ and $\varepsilon=10$ reduces both speaker-level and record-level MIA to near-random performance (AUC $\approx 0.52$--$0.54$, TPR@1\% $<0.02$), confirming its effectiveness against our attack. However, this formal privacy guarantee typically comes with a non-trivial trade-off in generation quality, which may be undesirable in practice. Due to space limits, we defer the full evaluation to Appendix~\ref{sec:defense_dp}.

\section{Related Works}
\noindent \textbf{Membership Inference Attacks on Generative Models.}
With the rapid growth of generative models, privacy auditing has become an increasingly important research topic~\cite{nasr2021adversary, xiang2025privacy, steinke2023privacy}. Membership inference attacks (MIAs), as a central auditing tool, have been extensively studied for generative models such as large language models and text-to-image generation models, moving from white-box settings toward more realistic black-box access. White-box attacks use internal model signals such as model parameters and gradients~\cite{pang2025white}, losses~\cite{hu2023membership, matsumoto2023membership}, or denoising trajectories~\cite{zhai2024membership,kong2023efficient}. In contrast, black-box attacks infer membership solely from observable input--output behavior, such as generated-sample similarity~\cite{pang2023black, du2025artistauditor, li2024towards} or token logits~\cite{shi2023detecting, zhang2024min}. Despite this progress, existing generative-model MIAs do not directly transfer to TTS. Modern TTS models use cross-modal dual conditioning on both text and reference audio, and their outputs are continuous, variable-length waveforms rather than discrete tokens or fixed-dimensional images. These differences make it challenging to construct memorization-triggering queries and to compare generated outputs with target records, motivating a TTS-specific black-box MIA framework.

\noindent \textbf{Speech Privacy.}
Speech carries rich sensitive information, including speaker identity, linguistic content, emotion, and other paralinguistic traits, which has motivated growing interest in speech privacy~\cite{lim2022overo, zhang2025speechguard, tomashenko2026third, tomashenko2025first, tseng2022membership}. In particular, MIAs have been investigated for speech understanding tasks, such as automatic speech recognition and automatic speaker recognition~\cite{shah2021evaluating, teixeira2024improving, chen2023slmia}. However, the privacy risks of generative TTS systems remain underexplored. The most relevant prior work~\cite{kong2023efficient} studies white-box membership inference for diffusion models, including limited experiments on early TTS systems conditioned only on text. Our work targets a fundamentally different setting: modern generative TTS systems that support text--audio dual conditioning for voice cloning, cover diverse model families beyond diffusion, and provide no access to internal states, gradients, or losses. 
As a result, the attacker must infer membership solely from synthesized waveforms, making black-box TTS membership inference substantially more challenging and largely unexplored.

\section{Conclusion}

We present the first black-box membership inference attack tailored to fine-tuned TTS models, auditing privacy leakage at both the speaker and record levels. We characterize the dual-conditioned query space and derive two criteria, scorable extent and memorization elicitation, to guide principled query design, under which recitation emerges as the strongest query. We further develop a representation engineering pipeline that combines multi-level speech representations with DTW-based temporal alignment to score continuous, variable-length outputs. Across three state-of-the-art models and two datasets, the attack consistently achieves an AUC above 0.8, and we further identify speech characteristics associated with disproportionate vulnerability to membership inference. The underlying principles of principled query design, multi-level representations, and temporal alignment can potentially generalize to generative models with multimodal conditioning and continuous outputs, highlighting the need for privacy-preserving fine-tuning as speech synthesis becomes increasingly personalized.

\section{Ethics Considerations}

This study follows responsible disclosure and ethical AI research principles. We conduct membership inference attacks only on publicly available datasets, including VCTK and British Dialect, and publicly released model checkpoints, including CosyVoice2, F5-TTS, and XTTS-v2. These attacks are conducted solely to evaluate and improve model privacy. We do not attempt to de-anonymize or identify any individuals associated with the data samples, and we do not access real user data or proprietary models without authorization. All experiments are performed in controlled environments using models that we fine-tune on public data. The goal of this work is to identify vulnerabilities in the fine-tuning pipeline, provide auditing tools against unauthorized use, and support the development of privacy-preserving techniques for speech generative models.

\bibliographystyle{IEEEtran}
\bibliography{refs}

\begin{thebibliography}{10}
\providecommand{\url}[1]{#1}
\csname url@samestyle\endcsname
\providecommand{\newblock}{\relax}
\providecommand{\bibinfo}[2]{#2}
\providecommand{\BIBentrySTDinterwordspacing}{\spaceskip=0pt\relax}
\providecommand{\BIBentryALTinterwordstretchfactor}{4}
\providecommand{\BIBentryALTinterwordspacing}{\spaceskip=\fontdimen2\font plus
\BIBentryALTinterwordstretchfactor\fontdimen3\font minus \fontdimen4\font\relax}
\providecommand{\BIBforeignlanguage}[2]{{%
\expandafter\ifx\csname l@#1\endcsname\relax
\typeout{** WARNING: IEEEtran.bst: No hyphenation pattern has been}%
\typeout{** loaded for the language `#1'. Using the pattern for}%
\typeout{** the default language instead.}%
\else
\language=\csname l@#1\endcsname
\fi
#2}}
\providecommand{\BIBdecl}{\relax}
\BIBdecl

\bibitem{wang2023neural}
C.~Wang, S.~Chen, Y.~Wu, Z.~Zhang, L.~Zhou, S.~Liu, Z.~Chen, Y.~Liu, H.~Wang, J.~Li \emph{et~al.}, ``Neural codec language models are zero-shot text to speech synthesizers,'' \emph{arXiv preprint arXiv:2301.02111}, 2023.

\bibitem{shen2023naturalspeech}
K.~Shen, Z.~Ju, X.~Tan, Y.~Liu, Y.~Leng, L.~He, T.~Qin, S.~Zhao, and J.~Bian, ``Naturalspeech 2: Latent diffusion models are natural and zero-shot speech and singing synthesizers,'' \emph{arXiv preprint arXiv:2304.09116}, 2023.

\bibitem{backstrom2025privacy}
T.~B{\"a}ckstr{\"o}m, ``Privacy in speech technology,'' \emph{Proceedings of the IEEE}, 2025.

\bibitem{tomashenko2022voiceprivacy}
N.~Tomashenko, X.~Wang, E.~Vincent, J.~Patino, B.~M.~L. Srivastava, P.-G. No{\'e}, A.~Nautsch, N.~Evans, J.~Yamagishi, B.~O’Brien \emph{et~al.}, ``The voiceprivacy 2020 challenge: Results and findings,'' \emph{Computer Speech \& Language}, vol.~74, p. 101362, 2022.

\bibitem{warren2024better}
K.~Warren, T.~Tucker, A.~Crowder, D.~Olszewski, A.~Lu, C.~Fedele, M.~Pasternak, S.~Layton, K.~Butler, C.~Gates \emph{et~al.}, ``" better be computer or i'm dumb": A large-scale evaluation of humans as audio deepfake detectors,'' in \emph{Proceedings of the 2024 on ACM SIGSAC Conference on Computer and Communications Security}, 2024, pp. 2696--2710.

\bibitem{gdpr2016}
\BIBentryALTinterwordspacing
{European Parliament and Council of the European Union}, ``Regulation (eu) 2016/679 of the european parliament and of the council of 27 april 2016,'' Official Journal of the European Union, L 119, 1--88, 2016. [Online]. Available: \url{https://eur-lex.europa.eu/eli/reg/2016/679/oj}
\BIBentrySTDinterwordspacing

\bibitem{euaiact2024}
\BIBentryALTinterwordspacing
------, ``Regulation (eu) 2024/1689 of the european parliament and of the council of 13 june 2024 laying down harmonised rules on artificial intelligence and amending regulations (ec) no 300/2008, (eu) no 167/2013, (eu) no 168/2013, (eu) 2018/858, (eu) 2018/1139 and (eu) 2019/2144 and directives 2014/90/eu, (eu) 2016/797 and (eu) 2020/1828,'' Official Journal of the European Union, L, 2024/1689, 2024. [Online]. Available: \url{https://eur-lex.europa.eu/eli/reg/2024/1689/oj}
\BIBentrySTDinterwordspacing

\bibitem{shokri2017membership}
R.~Shokri, M.~Stronati, C.~Song, and V.~Shmatikov, ``Membership inference attacks against machine learning models,'' in \emph{2017 IEEE symposium on security and privacy (SP)}.\hskip 1em plus 0.5em minus 0.4em\relax IEEE, 2017, pp. 3--18.

\bibitem{carlini2022membership}
N.~Carlini, S.~Chien, M.~Nasr, S.~Song, A.~Terzis, and F.~Tramer, ``Membership inference attacks from first principles,'' in \emph{2022 IEEE symposium on security and privacy (SP)}.\hskip 1em plus 0.5em minus 0.4em\relax IEEE, 2022, pp. 1897--1914.

\bibitem{hu2022membershipsok}
H.~Hu, Z.~Salcic, L.~Sun, G.~Dobbie, P.~S. Yu, and X.~Zhang, ``Membership inference attacks on machine learning: A survey,'' \emph{ACM Computing Surveys (CSUR)}, vol.~54, no. 11s, pp. 1--37, 2022.

\bibitem{salem2023sok}
A.~Salem, G.~Cherubin, D.~Evans, B.~K{\"o}pf, A.~Paverd, A.~Suri, S.~Tople, and S.~Zanella-B{\'e}guelin, ``Sok: Let the privacy games begin! a unified treatment of data inference privacy in machine learning,'' in \emph{2023 IEEE Symposium on Security and Privacy (SP)}.\hskip 1em plus 0.5em minus 0.4em\relax IEEE, 2023, pp. 327--345.

\bibitem{wu2022membership}
Y.~Wu, N.~Yu, Z.~Li, M.~Backes, and Y.~Zhang, ``Membership inference attacks against text-to-image generation models,'' \emph{arXiv preprint arXiv:2210.00968}, 2022.

\bibitem{pang2023black}
Y.~Pang and T.~Wang, ``Black-box membership inference attacks against fine-tuned diffusion models,'' \emph{arXiv preprint arXiv:2312.08207}, 2023.

\bibitem{duan2024membership}
M.~Duan, A.~Suri, N.~Mireshghallah, S.~Min, W.~Shi, L.~Zettlemoyer, Y.~Tsvetkov, Y.~Choi, D.~Evans, and H.~Hajishirzi, ``Do membership inference attacks work on large language models?'' \emph{arXiv preprint arXiv:2402.07841}, 2024.

\bibitem{shah2021evaluating}
M.~A. Shah, J.~Szurley, M.~Mueller, A.~Mouchtaris, and J.~Droppo, ``Evaluating the vulnerability of end-to-end automatic speech recognition models to membership inference attacks,'' in \emph{Proc. Interspeech 2021}, 2021, pp. 891--895.

\bibitem{chen2023slmia}
G.~Chen, Y.~Zhang, and F.~Song, ``Slmia-sr: Speaker-level membership inference attacks against speaker recognition systems,'' \emph{arXiv preprint arXiv:2309.07983}, 2023.

\bibitem{li2024towards}
J.~Li, J.~Dong, T.~He, and J.~Zhang, ``Towards black-box membership inference attack for diffusion models,'' \emph{arXiv preprint arXiv:2405.20771}, 2024.

\bibitem{stevens1937scale}
S.~S. Stevens, J.~Volkmann, and E.~B. Newman, ``A scale for the measurement of the psychological magnitude pitch,'' \emph{The journal of the acoustical society of america}, vol.~8, no.~3, pp. 185--190, 1937.

\bibitem{davis1980comparison}
S.~B. Davis and P.~Mermelstein, ``Comparison of parametric representations for monosyllabic word recognition in continuously spoken sentences,'' \emph{IEEE Transactions on Acoustics, Speech, and Signal Processing}, vol.~28, no.~4, pp. 357--366, 1980.

\bibitem{carlini2021extracting}
N.~Carlini, F.~Tramer, E.~Wallace, M.~Jagielski, A.~Herbert-Voss, K.~Lee, A.~Roberts, T.~Brown, D.~Song, U.~Erlingsson \emph{et~al.}, ``Extracting training data from large language models,'' in \emph{30th USENIX security symposium (USENIX Security 21)}, 2021, pp. 2633--2650.

\bibitem{shi2023detecting}
W.~Shi, A.~Ajith, M.~Xia, Y.~Huang, D.~Liu, T.~Blevins, D.~Chen, and L.~Zettlemoyer, ``Detecting pretraining data from large language models,'' \emph{arXiv preprint arXiv:2310.16789}, 2023.

\bibitem{duan2023diffusion}
J.~Duan, F.~Kong, S.~Wang, X.~Shi, and K.~Xu, ``Are diffusion models vulnerable to membership inference attacks?'' in \emph{International Conference on Machine Learning}.\hskip 1em plus 0.5em minus 0.4em\relax PMLR, 2023, pp. 8717--8730.

\bibitem{chen2022wavlm}
S.~Chen, C.~Wang, Z.~Chen, Y.~Wu, S.~Liu, Z.~Chen, J.~Li, N.~Kanda, T.~Yoshioka, X.~Xiao \emph{et~al.}, ``Wavlm: Large-scale self-supervised pre-training for full stack speech processing,'' \emph{IEEE Journal of Selected Topics in Signal Processing}, vol.~16, no.~6, pp. 1505--1518, 2022.

\bibitem{sakoe2003dynamic}
H.~Sakoe and S.~Chiba, ``Dynamic programming algorithm optimization for spoken word recognition,'' \emph{IEEE transactions on acoustics, speech, and signal processing}, vol.~26, no.~1, pp. 43--49, 2003.

\bibitem{du2024cosyvoice}
Z.~Du, Y.~Wang, Q.~Chen, X.~Shi, X.~Lv, T.~Zhao, Z.~Gao, Y.~Yang, C.~Gao, H.~Wang \emph{et~al.}, ``Cosyvoice 2: Scalable streaming speech synthesis with large language models,'' \emph{arXiv preprint arXiv:2412.10117}, 2024.

\bibitem{casanova2024xtts}
E.~Casanova, K.~Davis, E.~G{\"o}lge, G.~G{\"o}knar, I.~Gulea, L.~Hart, A.~Aljafari, J.~Meyer, R.~Morais, S.~Olayemi \emph{et~al.}, ``Xtts: a massively multilingual zero-shot text-to-speech model,'' \emph{arXiv preprint arXiv:2406.04904}, 2024.

\bibitem{chen2025f5}
Y.~Chen, Z.~Niu, Z.~Ma, K.~Deng, C.~Wang, J.~JianZhao, K.~Yu, and X.~Chen, ``F5-tts: A fairytaler that fakes fluent and faithful speech with flow matching,'' in \emph{Proceedings of the 63rd Annual Meeting of the Association for Computational Linguistics (Volume 1: Long Papers)}, 2025, pp. 6255--6271.

\bibitem{yamagishi2019cstr}
J.~Yamagishi, C.~Veaux, and K.~MacDonald, ``Cstr vctk corpus: English multi-speaker corpus for cstr voice cloning toolkit (version 0.92),'' \emph{The Rainbow Passage which the speakers read out can be found in the International Dialects of English Archive:(http://web. ku. edu/\~{} idea/readings/rainbow. htm).}, 2019.

\bibitem{demirsahin2020open}
I.~Demirsahin, O.~Kjartansson, A.~Gutkin, and C.~Rivera, ``Open-source multi-speaker corpora of the english accents in the british isles,'' in \emph{Proceedings of the twelfth language resources and evaluation conference}, 2020, pp. 6532--6541.

\bibitem{wood2018does}
S.~G. Wood, J.~H. Moxley, E.~L. Tighe, and R.~K. Wagner, ``Does use of text-to-speech and related read-aloud tools improve reading comprehension for students with reading disabilities? a meta-analysis,'' \emph{Journal of learning disabilities}, vol.~51, no.~1, pp. 73--84, 2018.

\bibitem{tan2021survey}
X.~Tan, T.~Qin, F.~Soong, and T.-Y. Liu, ``A survey on neural speech synthesis,'' \emph{arXiv preprint arXiv:2106.15561}, 2021.

\bibitem{kathiria2024assistive}
P.~Kathiria, S.~H. Mankad, J.~Patel, M.~Kapadia, and N.~Lakdawala, ``Assistive systems for visually impaired people: A survey on current requirements and advancements,'' \emph{Neurocomputing}, vol. 606, p. 128284, 2024.

\bibitem{shen2018natural}
J.~Shen, R.~Pang, R.~J. Weiss, M.~Schuster, N.~Jaitly, Z.~Yang, Z.~Chen, Y.~Zhang, Y.~Wang, R.~Skerrv-Ryan \emph{et~al.}, ``Natural tts synthesis by conditioning wavenet on mel spectrogram predictions,'' in \emph{2018 IEEE international conference on acoustics, speech and signal processing (ICASSP)}.\hskip 1em plus 0.5em minus 0.4em\relax IEEE, 2018, pp. 4779--4783.

\bibitem{kim2021conditional}
J.~Kim, J.~Kong, and J.~Son, ``Conditional variational autoencoder with adversarial learning for end-to-end text-to-speech,'' in \emph{International conference on machine learning}.\hskip 1em plus 0.5em minus 0.4em\relax PMLR, 2021, pp. 5530--5540.

\bibitem{xie2025towards}
T.~Xie, Y.~Rong, P.~Zhang, W.~Wang, and L.~Liu, ``Towards controllable speech synthesis in the era of large language models: A systematic survey,'' in \emph{Proceedings of the 2025 Conference on Empirical Methods in Natural Language Processing}, 2025, pp. 764--791.

\bibitem{azzuni2025voice}
H.~Azzuni and A.~E. Saddik, ``Voice cloning: Comprehensive survey,'' \emph{arXiv preprint arXiv:2505.00579}, 2025.

\bibitem{hu2026qwen3}
H.~Hu, X.~Zhu, T.~He, D.~Guo, B.~Zhang, X.~Wang, Z.~Guo, Z.~Jiang, H.~Hao, Z.~Guo \emph{et~al.}, ``Qwen3-tts technical report,'' \emph{arXiv preprint arXiv:2601.15621}, 2026.

\bibitem{le2023voicebox}
M.~Le, A.~Vyas, B.~Shi, B.~Karrer, L.~Sari, R.~Moritz, M.~Williamson, V.~Manohar, Y.~Adi, J.~Mahadeokar \emph{et~al.}, ``Voicebox: Text-guided multilingual universal speech generation at scale,'' \emph{Advances in neural information processing systems}, vol.~36, pp. 14\,005--14\,034, 2023.

\bibitem{eskimez2024e2}
S.~E. Eskimez, X.~Wang, M.~Thakker, C.~Li, C.-H. Tsai, Z.~Xiao, H.~Yang, Z.~Zhu, M.~Tang, X.~Tan \emph{et~al.}, ``E2 tts: Embarrassingly easy fully non-autoregressive zero-shot tts,'' in \emph{2024 IEEE spoken language technology workshop (SLT)}.\hskip 1em plus 0.5em minus 0.4em\relax IEEE, 2024, pp. 682--689.

\bibitem{du2024cosyvoice1}
Z.~Du, Q.~Chen, S.~Zhang, K.~Hu, H.~Lu, Y.~Yang, H.~Hu, S.~Zheng, Y.~Gu, Z.~Ma \emph{et~al.}, ``Cosyvoice: A scalable multilingual zero-shot text-to-speech synthesizer based on supervised semantic tokens,'' \emph{arXiv preprint arXiv:2407.05407}, 2024.

\bibitem{chatterboxtts2025}
{Resemble AI}, ``{Chatterbox-TTS},'' \url{https://github.com/resemble-ai/chatterbox}, 2025, gitHub repository.

\bibitem{zhai2024membership}
S.~Zhai, H.~Chen, Y.~Dong, J.~Li, Q.~Shen, Y.~Gao, H.~Su, and Y.~Liu, ``Membership inference on text-to-image diffusion models via conditional likelihood discrepancy,'' \emph{Advances in Neural Information Processing Systems}, vol.~37, pp. 74\,122--74\,146, 2024.

\bibitem{carlini2023extracting}
N.~Carlini, J.~Hayes, M.~Nasr, M.~Jagielski, V.~Sehwag, F.~Tramer, B.~Balle, D.~Ippolito, and E.~Wallace, ``Extracting training data from diffusion models,'' in \emph{32nd USENIX security symposium (USENIX Security 23)}, 2023, pp. 5253--5270.

\bibitem{nasr2021adversary}
M.~Nasr, S.~Songi, A.~Thakurta, N.~Papernot, and N.~Carlin, ``Adversary instantiation: Lower bounds for differentially private machine learning,'' in \emph{2021 IEEE Symposium on security and privacy (SP)}.\hskip 1em plus 0.5em minus 0.4em\relax IEEE, 2021, pp. 866--882.

\bibitem{ippolito2022preventing}
D.~Ippolito, F.~Tram{\`e}r, M.~Nasr, C.~Zhang, M.~Jagielski, K.~Lee, C.~A. Choquette-Choo, and N.~Carlini, ``Preventing verbatim memorization in language models gives a false sense of privacy,'' \emph{arXiv preprint arXiv:2210.17546}, 2022.

\bibitem{yeom2018privacy}
S.~Yeom, I.~Giacomelli, M.~Fredrikson, and S.~Jha, ``Privacy risk in machine learning: Analyzing the connection to overfitting,'' in \emph{2018 IEEE 31st Computer Security Foundations Symposium (CSF)}.\hskip 1em plus 0.5em minus 0.4em\relax IEEE, 2018, pp. 268--282.

\bibitem{chen2020ganleaks}
D.~Chen, N.~Yu, Y.~Zhang, and M.~Fritz, ``{GAN-Leaks}: A taxonomy of membership inference attacks against generative models,'' in \emph{Proceedings of the 2020 ACM SIGSAC Conference on Computer and Communications Security (CCS)}, 2020, pp. 343--362.

\bibitem{zhu2026omnivoice}
H.~Zhu, L.~Ye, W.~Kang, Z.~Yao, L.~Guo, F.~Kuang, Z.~Han, W.~Zhuang, L.~Lin, and D.~Povey, ``Omnivoice: Towards omnilingual zero-shot text-to-speech with diffusion language models,'' \emph{arXiv preprint arXiv:2604.00688}, 2026.

\bibitem{carlini2022quantifying}
N.~Carlini, D.~Ippolito, M.~Jagielski, K.~Lee, F.~Tramer, and C.~Zhang, ``Quantifying memorization across neural language models,'' \emph{arXiv preprint arXiv:2202.07646}, 2022.

\bibitem{ren2022revisiting}
Y.~Ren, X.~Tan, T.~Qin, Z.~Zhao, and T.-Y. Liu, ``Revisiting over-smoothness in text to speech,'' in \emph{Proceedings of the 60th Annual Meeting of the Association for Computational Linguistics (Volume 1: Long Papers)}, 2022, pp. 8197--8213.

\bibitem{zhou2022content}
Y.~Zhou, C.~Song, X.~Li, L.~Zhang, Z.~Wu, Y.~Bian, D.~Su, and H.~Meng, ``Content-dependent fine-grained speaker embedding for zero-shot speaker adaptation in text-to-speech synthesis,'' \emph{arXiv preprint arXiv:2204.00990}, 2022.

\bibitem{snyder2018x}
D.~Snyder, D.~Garcia-Romero, G.~Sell, D.~Povey, and S.~Khudanpur, ``X-vectors: Robust dnn embeddings for speaker recognition,'' in \emph{2018 IEEE international conference on acoustics, speech and signal processing (ICASSP)}.\hskip 1em plus 0.5em minus 0.4em\relax IEEE, 2018, pp. 5329--5333.

\bibitem{DesplanquesTD20}
\BIBentryALTinterwordspacing
B.~Desplanques, J.~Thienpondt, and K.~Demuynck, ``{ECAPA-TDNN:} emphasized channel attention, propagation and aggregation in {TDNN} based speaker verification,'' in \emph{21st Annual Conference of the International Speech Communication Association, Interspeech 2020, Virtual Event, Shanghai, China, October 25-29, 2020}, H.~Meng, B.~Xu, and T.~F. Zheng, Eds.\hskip 1em plus 0.5em minus 0.4em\relax {ISCA}, 2020, pp. 3830--3834. [Online]. Available: \url{https://doi.org/10.21437/Interspeech.2020-2650}
\BIBentrySTDinterwordspacing

\bibitem{chen2022large}
Z.~Chen, S.~Chen, Y.~Wu, Y.~Qian, C.~Wang, S.~Liu, Y.~Qian, and M.~Zeng, ``Large-scale self-supervised speech representation learning for automatic speaker verification,'' in \emph{ICASSP 2022-2022 IEEE International Conference on Acoustics, Speech and Signal Processing (ICASSP)}.\hskip 1em plus 0.5em minus 0.4em\relax IEEE, 2022, pp. 6147--6151.

\bibitem{liu2020text}
K.~Liu and H.~Zhou, ``Text-independent speaker verification with adversarial learning on short utterances,'' in \emph{ICASSP 2020-2020 IEEE International Conference on Acoustics, Speech and Signal Processing (ICASSP)}.\hskip 1em plus 0.5em minus 0.4em\relax IEEE, 2020, pp. 6569--6573.

\bibitem{he2024emilia}
H.~He, Z.~Shang, C.~Wang, X.~Li, Y.~Gu, H.~Hua, L.~Liu, C.~Yang, J.~Li, P.~Shi \emph{et~al.}, ``Emilia: An extensive, multilingual, and diverse speech dataset for large-scale speech generation,'' in \emph{2024 IEEE Spoken Language Technology Workshop (SLT)}.\hskip 1em plus 0.5em minus 0.4em\relax IEEE, 2024, pp. 885--890.

\bibitem{koizumi2023libritts}
Y.~Koizumi, H.~Zen, S.~Karita, Y.~Ding, K.~Yatabe, N.~Morioka, M.~Bacchiani, Y.~Zhang, W.~Han, and A.~Bapna, ``Libritts-r: A restored multi-speaker text-to-speech corpus,'' in \emph{Proc. Interspeech 2023}, 2023, pp. 5496--5500.

\bibitem{kahn2020libri}
J.~Kahn, M.~Riviere, W.~Zheng, E.~Kharitonov, Q.~Xu, P.-E. Mazar{\'e}, J.~Karadayi, V.~Liptchinsky, R.~Collobert, C.~Fuegen \emph{et~al.}, ``Libri-light: A benchmark for asr with limited or no supervision,'' in \emph{ICASSP 2020-2020 IEEE International Conference on Acoustics, Speech and Signal Processing (ICASSP)}.\hskip 1em plus 0.5em minus 0.4em\relax IEEE, 2020, pp. 7669--7673.

\bibitem{childers1991gender}
D.~G. Childers and K.~Wu, ``Gender recognition from speech. part ii: Fine analysis,'' \emph{The Journal of the Acoustical society of America}, vol.~90, no.~4, pp. 1841--1856, 1991.

\bibitem{bak2023avocodo}
T.~Bak, J.~Lee, H.~Bae, J.~Yang, J.-S. Bae, and Y.-S. Joo, ``Avocodo: Generative adversarial network for artifact-free vocoder,'' in \emph{Proceedings of the AAAI Conference on Artificial Intelligence}, vol.~37, no.~11, 2023, pp. 12\,562--12\,570.

\bibitem{ju2024naturalspeech}
Z.~Ju, Y.~Wang, K.~Shen, X.~Tan, D.~Xin, D.~Yang, Y.~Liu, Y.~Leng, K.~Song, S.~Tang \emph{et~al.}, ``Naturalspeech 3: Zero-shot speech synthesis with factorized codec and diffusion models,'' \emph{arXiv preprint arXiv:2403.03100}, 2024.

\bibitem{kuanyv2v}
M.~Ku, C.~Wei, W.~Ren, H.~Yang, and W.~Chen, ``Anyv2v: A tuning-free framework for any video-to-video editing tasks,'' \emph{Transactions on Machine Learning Research}.

\bibitem{xu2025qwen3}
J.~Xu, Z.~Guo, H.~Hu, Y.~Chu, X.~Wang, J.~He, Y.~Wang, X.~Shi, T.~He, X.~Zhu \emph{et~al.}, ``Qwen3-omni technical report,'' \emph{arXiv preprint arXiv:2509.17765}, 2025.

\bibitem{xie2024mini}
Z.~Xie and C.~Wu, ``Mini-omni2: Towards open-source gpt-4o with vision, speech and duplex capabilities,'' \emph{arXiv preprint arXiv:2410.11190}, 2024.

\bibitem{abadi2016deep}
M.~Abadi, A.~Chu, I.~Goodfellow, H.~B. McMahan, I.~Mironov, K.~Talwar, and L.~Zhang, ``Deep learning with differential privacy,'' in \emph{Proceedings of the 2016 ACM SIGSAC conference on computer and communications security}, 2016, pp. 308--318.

\bibitem{xiang2025privacy}
Z.~Xiang, T.~Wang, and D.~Wang, ``Privacy audit as bits transmission:(im) possibilities for audit by one run,'' in \emph{34th USENIX Security Symposium (USENIX Security 25)}, 2025, pp. 2693--2711.

\bibitem{steinke2023privacy}
T.~Steinke, M.~Nasr, and M.~Jagielski, ``Privacy auditing with one (1) training run,'' \emph{Advances in Neural Information Processing Systems}, vol.~36, pp. 49\,268--49\,280, 2023.

\bibitem{pang2025white}
Y.~Pang, T.~Wang, X.~Kang, M.~Huai, and Y.~Zhang, ``White-box membership inference attacks against diffusion models,'' \emph{Proceedings on Privacy Enhancing Technologies}, vol. 2025, no.~2, 2025.

\bibitem{hu2023membership}
H.~Hu and J.~Pang, ``Membership inference of diffusion models,'' \emph{arXiv preprint arXiv:2301.09956}, 2023.

\bibitem{matsumoto2023membership}
T.~Matsumoto, T.~Miura, and N.~Yanai, ``Membership inference attacks against diffusion models,'' in \emph{2023 IEEE Security and Privacy Workshops (SPW)}.\hskip 1em plus 0.5em minus 0.4em\relax IEEE, 2023, pp. 77--83.

\bibitem{kong2023efficient}
F.~Kong, J.~Duan, R.~Ma, H.~Shen, X.~Zhu, X.~Shi, and K.~Xu, ``An efficient membership inference attack for the diffusion model by proximal initialization,'' \emph{arXiv preprint arXiv:2305.18355}, 2023.

\bibitem{du2025artistauditor}
L.~Du, Z.~Zhu, M.~Chen, Z.~Su, S.~Ji, P.~Cheng, J.~Chen, and Z.~Zhang, ``Artistauditor: Auditing artist style pirate in text-to-image generation models,'' in \emph{Proceedings of the ACM on Web Conference 2025}, 2025, pp. 2500--2513.

\bibitem{zhang2024min}
J.~Zhang, J.~Sun, E.~Yeats, Y.~Ouyang, M.~Kuo, J.~Zhang, H.~F. Yang, and H.~Li, ``Min-k\%++: Improved baseline for detecting pre-training data from large language models,'' \emph{arXiv preprint arXiv:2404.02936}, 2024.

\bibitem{lim2022overo}
J.~Lim, K.~Kim, H.~Yu, and S.-B. Lee, ``Overo: Sharing private audio recordings,'' in \emph{Proceedings of the 2022 ACM SIGSAC Conference on Computer and Communications Security}, 2022, pp. 1933--1946.

\bibitem{zhang2025speechguard}
J.~Zhang, S.~Liu, J.~Hou, Z.~Wang, H.~Yu, and X.-Y. Li, ``$\{$SpeechGuard$\}$: Recoverable and customizable speech privacy protection,'' in \emph{34th USENIX Security Symposium (USENIX Security 25)}, 2025, pp. 5931--5948.

\bibitem{tomashenko2026third}
N.~Tomashenko, X.~Miao, P.~Champion, S.~Meyer, M.~Panariello, X.~Wang, N.~Evans, E.~Vincent, J.~Yamagishi, and M.~Todisco, ``The third voiceprivacy challenge: Preserving emotional expressiveness and linguistic content in voice anonymization,'' \emph{arXiv preprint arXiv:2601.11846}, 2026.

\bibitem{tomashenko2025first}
N.~Tomashenko, X.~Miao, E.~Vincent, and J.~Yamagishi, ``The first voiceprivacy attacker challenge,'' in \emph{ICASSP 2025-2025 IEEE International Conference on Acoustics, Speech and Signal Processing (ICASSP)}.\hskip 1em plus 0.5em minus 0.4em\relax IEEE, 2025, pp. 1--2.

\bibitem{tseng2022membership}
W.-C. Tseng, W.-T. Kao, and H.-y. Lee, ``Membership inference attacks against self-supervised speech models,'' in \emph{Proc. Interspeech 2022}, 2022, pp. 5040--5044.

\bibitem{teixeira2024improving}
F.~Teixeira, K.~Pizzi, R.~Olivier, A.~Abad, B.~Raj, and I.~Trancoso, ``Improving membership inference in asr model auditing with perturbed loss features,'' \emph{arXiv preprint arXiv:2405.01207}, 2024.

\bibitem{weide1998cmu}
R.~L. Weide, ``The {CMU} pronouncing dictionary,'' \url{http://www.speech.cs.cmu.edu/cgi-bin/cmudict}, 1998.

\end{thebibliography}

\appendix

\begin{algorithm}[htb]
\caption{Generalized Membership Inference Auditing for TTS}
\label{alg:mia_tts}

\begin{algorithmic}[1]

\Statex \textbf{Input:} Target generative model $\mathcal{F}$, target data $x$, number of queries $m$, decision threshold $\tau$
\Statex \textbf{Output:} Membership decision $\hat{y} \in \{0,1\}$
\Statex

\Statex \textit{\textbf{\# Phase 1: Query Formulation \& Execution}}

\State $q \gets \textsc{RecitationQuery}(x)$
\Statex \textcolor{blue}{$\triangleright$ Our Recitation query $G_Q$}

\State $\mathcal{S}_{\text{gen}} \gets \emptyset$

\For{$j = 1$ to $m$}
    \State $s_{\text{gen}}^{(j)} \gets \mathcal{F}(q)$
    \State $\mathcal{S}_{\text{gen}} \gets \mathcal{S}_{\text{gen}} \cup \{s_{\text{gen}}^{(j)}\}$
\EndFor
\Statex

\Statex \textit{\textbf{\# Phase 2: Representation Engineering ($\phi_{\text{spk}}$ / $\phi_{\text{rec}}$)}}

\State $\mathcal{H}_{\text{feat}} \gets \textsc{ExtractRepresentation}(x,\, \mathcal{S}_{\text{gen}})$
\Statex \textcolor{blue}{$\triangleright$ Representation extraction tailored to different attack goals}

\State $\mathcal{H}_{\text{align}} \gets \textsc{AlignRepresentation}(\mathcal{H}_{\text{feat}})$
\Statex \textcolor{blue}{$\triangleright$ Temporal alignments for variable-length waveforms}
\Statex

\Statex \textit{\textbf{\# Phase 3: Scoring \& Decision}}

\State $\mathcal{M}(x) \gets \textsc{ComputeScore}(\mathcal{H}_{\text{align}})$
\Statex \textcolor{blue}{$\triangleright$ Calculate aggregated score}

\State $\hat{y} \gets \mathbb{I}[\mathcal{M}(x) > \tau]$
\Statex \textcolor{blue}{$\triangleright$ Threshold decision}

\State \Return $\hat{y}$

\end{algorithmic}
\end{algorithm}

\subsection{Other Speech Characteristics Affecting Record-Level MIA}
\label{sec:appendix_record_mia_factors}

\noindent\textbf{Higher Speech Energy Increases Record-Level Vulnerability}
\begin{table}[t]
\centering
\caption{Record-level MIA performance across speech energy bins; higher energy recordings are more vulnerable.}
\label{tab:speech_energy_bins}
\setlength{\tabcolsep}{8pt}
\begin{tabular}{lccc}
\toprule
\textbf{Bin} & \textbf{AUC} & \textbf{ACC} & \textbf{TPR@1\%FPR} \\
\midrule
$<$0.02            & 0.845 & 0.713 & 0.242 \\
0.02--0.04         & 0.865 & 0.763 & 0.221 \\
0.04--0.06         & 0.901 & 0.814 & 0.308 \\
0.06--0.08         & 0.922 & 0.838 & 0.358 \\
$>$0.08            & 0.915 & 0.824 & 0.330 \\
\bottomrule
\end{tabular}
\end{table}

Speech energy measures the average RMS amplitude over voiced regions of an utterance. Table~\ref{tab:speech_energy_bins} shows a clear increasing relationship between speech energy and attack performance. AUC rises from 0.845 in the lowest bin ($<$0.02) to 0.922 in the 0.06--0.08 bin, and TPR@1\%FPR increases from 0.242 to 0.358 over the same range. We attribute this trend to the fact that higher-energy recordings contain richer harmonic and formant structure, providing more discriminative acoustic evidence for the DTW-based scoring pipeline to detect memorization.

\subsection{Effect of Multi-Query Statistics on Record-Level MIA}
\label{sec:appendix_mean_var}
\begin{table}[t]
\centering
\caption{Effect of different multi-query statistics on record-level MIA performance. The mean alone provides a modest gain over a single query, variance alone is weak, and combining mean with variance yields the best results.}
\label{tab:mean_meanvar}
\setlength{\tabcolsep}{8pt}
\begin{tabular}{lccc}
\toprule
\textbf{Method} & \textbf{AUC} & \textbf{ACC} & \textbf{TPR@1\%FPR} \\
\midrule
Single Query & 0.863 & 0.780 & 0.226 \\
Mean         & 0.869 & 0.784 & 0.265 \\
Var          & 0.690 & 0.644 & 0.033 \\
Mean+Var     & 0.896 & 0.809 & 0.281 \\
\bottomrule
\end{tabular}
\end{table}

Table~\ref{tab:mean_meanvar} ablates the contribution of different multi-query aggregation statistics to record-level MIA performance. A single query already provides a strong baseline, with an AUC of 0.863 and a TPR@1\%FPR of 0.226. Aggregating the mean score across multiple queries yields a modest but consistent improvement, increasing the AUC to 0.869 and the TPR@1\%FPR to 0.265, likely because averaging reduces noise from stochastic generation. In contrast, using only the score variance is substantially weaker, with an AUC of 0.690 and a TPR@1\%FPR of 0.033, suggesting that variance alone is not a reliable membership signal. Combining the mean and variance gives the best overall performance, reaching an AUC of 0.896 and a TPR@1\%FPR of 0.281. We attribute this improvement to the complementary membership signals captured by the two statistics: member samples tend to have higher similarity scores and lower variation across repeated queries.

\subsection{Query Configurations Used in the Experiments}
\label{app:query_config}

This appendix details the concrete configuration of the five queries compared in Section~\ref{sec:ablation_query}. All queries are issued through the zero-shot inference API of the target model, which takes a synthesis text and a reference speech.

\noindent\textbf{Recitation.} The query supplies the complete record, using the full recording as the reference speech and its transcript as the synthesis text.

\noindent\textbf{Continuation.} Each recording is cut at its temporal midpoint ($\rho=0.5$), with the cut placed at the boundary of the word spanning the midpoint according to word-level forced alignment. The first half and its transcript serve as the reference, and the model is asked to synthesize the remaining text.

\noindent\textbf{Partial reference.} A word-aligned window covering half of the recording ($\rho=0.5$) is taken at a random position and serves as the reference speech; the synthesis text is the transcript of the same window, so the model is asked to reproduce the clip.

\noindent\textbf{Audio perturbation.} The reference speech is the full recording corrupted with additive white Gaussian noise, with the perturbation strength $\sigma$ set to an SNR of 20\,dB and an independent noise realization per record; the synthesis text is the full transcript.

\noindent\textbf{Text perturbation.} A fraction $\delta=0.5$ of the content words in the synthesis text is replaced with near-homophonic alternatives, selected from the CMU Pronouncing Dictionary~\cite{weide1998cmu} by a feature-weighted phoneme edit distance; stopwords, very short words, proper nouns, and same-stem inflections are excluded from replacement. The replacement is fixed across generation seeds, and the reference speech is the full recording.

\subsection{Qualitative Example of Memorized Details in Reconstruction}
\label{app:recon_example}
\begin{figure*}[t]
  \centering
  \includegraphics[width=0.9\linewidth]{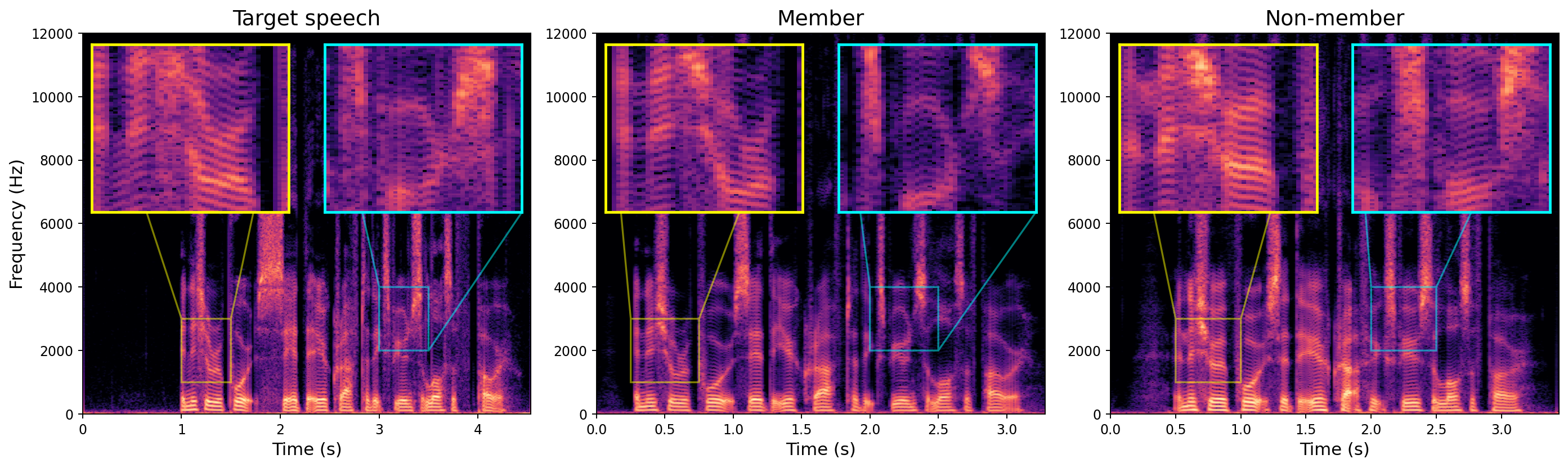}
  \caption{Spectrograms of a target speech (left) and its reconstructions under the recitation query by a model that was trained on the recording (middle, member) and by a model trained with the same recipe but without it (right, non-member). The member reconstruction reproduces the harmonic and formant structure closely, whereas the non-member reconstruction renders the same content with a smoother, generic pattern that lacks detail.}
  \label{fig:recon_example}
\end{figure*}

Figure~\ref{fig:recon_example} illustrates the mechanism behind criterion C2 in Section~\ref{sec:design_principle} on a single record, which is reconstructed by two models trained with the same recipe, one with the record in its training data and one without. Both models receive the same recitation query and produce intelligible speech in a voice close to that of the target, so the two reconstructions agree with the target in overall voice characteristics and content. They differ in the fine detail. In the enlarged regions, the model that has seen the recording reproduces the harmonic and formant structure of the target closely, whereas the model that has not seen it renders the same content with a smoother, generic pattern that lacks such detail. Membership therefore manifests in how the content is realized rather than in what is produced, which is the detail the record-level comparison in Section~\ref{sec:representation_learning} is designed to capture.

\subsection{Effect of Reference Completeness on MIA Scores}
\label{app:ref_length}
\begin{figure}[t]
  \centering
  \includegraphics[width=\linewidth]{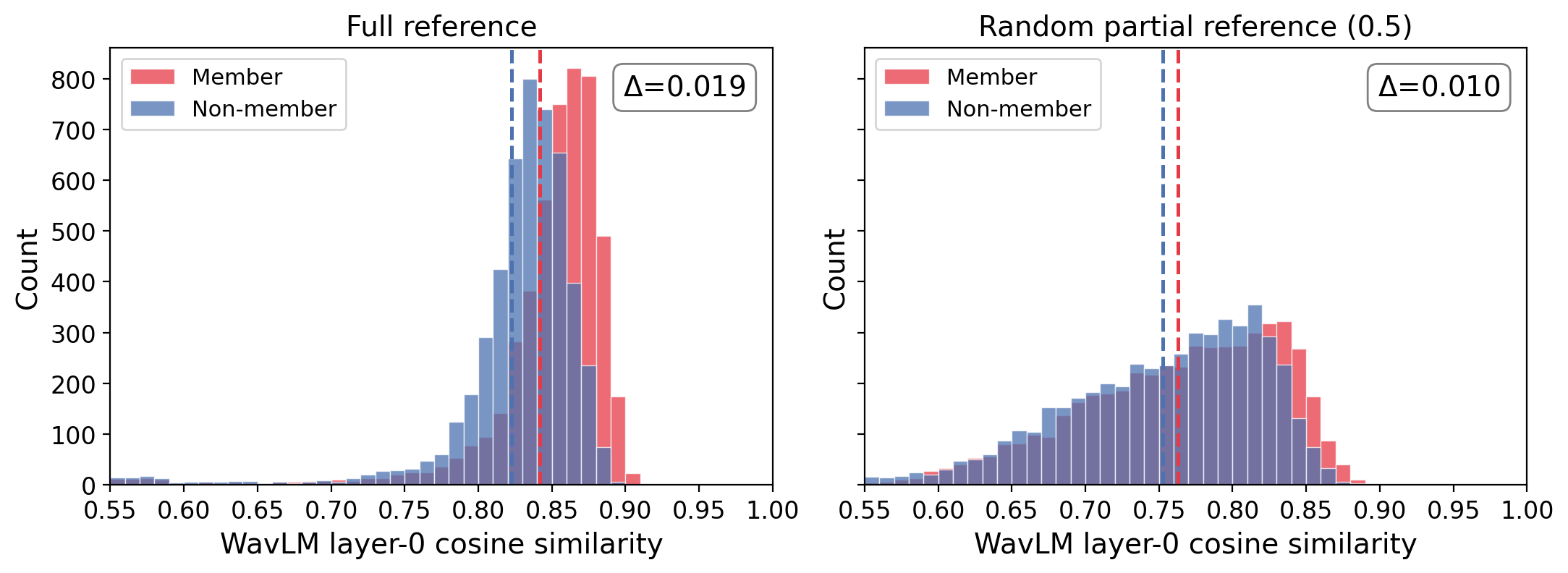}
  \caption{Reconstruction scores of members and non-members when the synthesis text is fixed to $t^{*}$ and the reference speech is either the full $s^{*}$ (left) or a random window covering half of it (right). Dashed lines mark the mean of each population and $\Delta$ denotes the gap between the two means.}
  \label{fig:ref_length}
\end{figure}

This appendix examines how the completeness of the reference speech affects the reconstruction scores of members and non-members, isolating criterion C2 in Section~\ref{sec:design_principle}. We fix the synthesis text to the full transcript $t^{*}$, so that the response remains fully comparable with the target under C1, and vary only the reference speech: the full $s^{*}$ versus a random window covering half of it, on the same records and generation seeds. To avoid any learned component, the score is the mean frame-wise cosine similarity between WavLM layer-0 features of the generated and the target speech.

Figure~\ref{fig:ref_length} shows the score distributions. Supplying the full reference raises the scores of both populations: the distributions of members and non-members both shift toward higher similarity and become more concentrated than under the partial reference. The increase, however, is not uniform. The member distribution moves further than the non-member distribution, and the gap between the two means roughly doubles, from $\Delta=0.010$ under the partial reference to $\Delta=0.019$ under the full reference. A more complete reference thus constrains the output for every record but moves the output of a memorized record closer to its recorded realization, which widens the separation between the two populations rather than shrinking it, as predicted by C2.

\subsection{Effect of Training Steps on Attack Success}
\label{app:training_steps}
\begin{figure}[t]
  \centering
  \subfloat[Speaker-level MIA.]{%
        \includegraphics[width=0.49\linewidth]{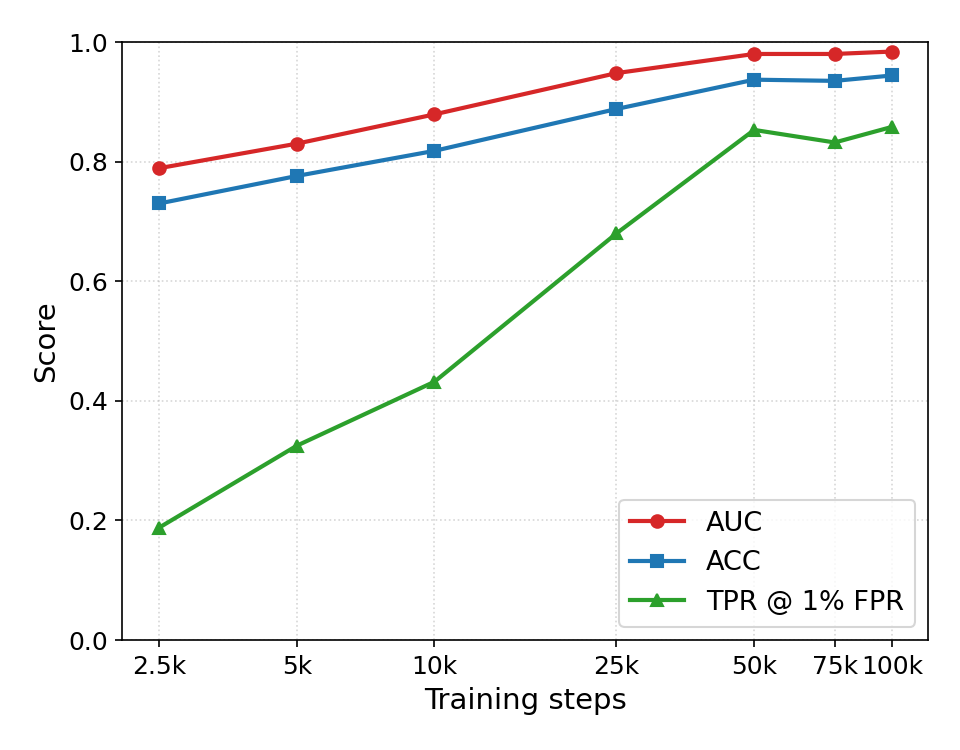}%
        \label{fig:epoch_sweep_spk}}
  \hfill
  \subfloat[Record-level MIA.]{%
        \includegraphics[width=0.49\linewidth]{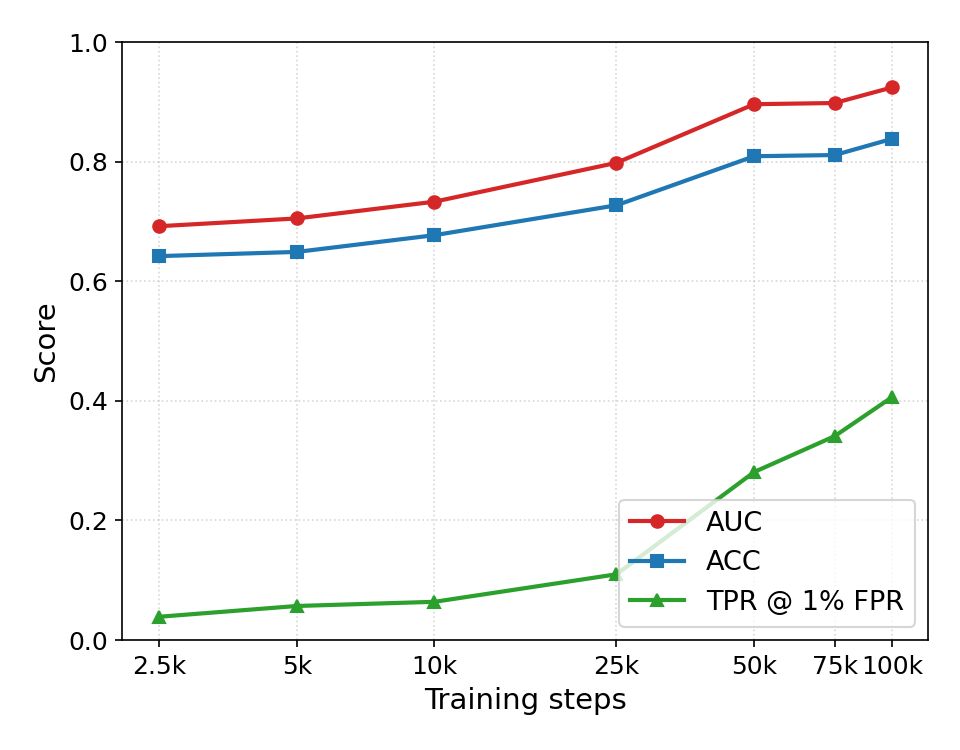}%
        \label{fig:epoch_sweep_utt}}
  \caption{Attack performance versus the number of fine-tuning steps for CosyVoice2 on VCTK with the recitation query.}
  \label{fig:epoch_sweep}
\end{figure}

Figure~\ref{fig:epoch_sweep} shows how attack performance varies with the number of fine-tuning steps for CosyVoice2 on VCTK. Attack success increases monotonically with training at both levels, as additional fine-tuning steps let the model fit the fine-tuning records more closely and thereby strengthen the membership signal~\cite{yeom2018privacy}. Speaker-level MIA saturates at around 50k steps, where the AUC reaches 0.980 with a TPR@1\%FPR of 0.853 and further training yields only marginal gains (AUC 0.984 at 100k steps). Record-level MIA gains most of its performance by 50k steps as well, rising from an AUC of 0.692 at 2.5k steps to 0.896 at 50k steps, and continues to improve more slowly thereafter, reaching an AUC of 0.924 and a TPR@1\%FPR of 0.406 at 100k steps.

Notably, the attacks remain effective even after few fine-tuning steps. At 2.5k steps, speaker-level MIA already attains an AUC of 0.789 with a TPR@1\%FPR of 0.188, and record-level MIA attains an AUC of 0.692 with a TPR@1\%FPR of 0.039, both well above the random-guess baseline of 0.5. Membership leakage therefore emerges early in fine-tuning rather than only after extensive training.

\subsection{Defense}
\label{sec:appendix_defense}
\subsubsection{Early Stop}
\label{sec:defense_early_stop}
\begin{table}[t]
\centering
\caption{Attack performance under early stopping defense for speaker-level and record-level MIA.}
\label{tab:early_stop_defense}
\setlength{\tabcolsep}{6pt}
\begin{tabular}{@{}lccc@{}}
\toprule
Level & AUC & ACC & TPR@1\%FPR \\
\midrule
Speaker & $0.900$ & $0.836$ & $0.543$ \\
\midrule
Record  & $0.748$ & $0.689$ & $0.036$ \\
\bottomrule
\end{tabular}
\end{table}

As is shown in Table~\ref{tab:early_stop_defense}, early stopping reduces record-level MIA substantially (AUC drops to 0.748, TPR@1\%FPR to 0.036), as the model has less opportunity to overfit fine-grained content details. However, speaker-level MIA remains strong (AUC 0.900, TPR@1\%FPR 0.543), indicating that speaker identity is memorized early in training and is hard to be effectively suppressed by this approach.

\subsubsection{Input Perturbation}
\label{sec:defense_input_perturb}
\begin{table}[t]
\centering
\caption{Attack performance under perturbation-based defenses.}
\label{tab:perturb_defense}
\setlength{\tabcolsep}{6pt}
\begin{tabular}{@{}lllccc@{}}
\toprule
Perturb & Level & Param & AUC & ACC & TPR@1\%FPR \\
\midrule
\multirow{4}{*}{Noise}
  & \multirow{2}{*}{Speaker} & 40 dB & $0.970$ & $0.925$ & $0.814$ \\
  &                          & 25 dB & $0.923$ & $0.878$ & $0.559$ \\
  \cmidrule{2-6}
  & \multirow{2}{*}{Record}  & 40 dB & $0.838$ & $0.761$ & $0.157$ \\
  &                          & 25 dB & $0.808$ & $0.732$ & $0.126$ \\
\midrule
\multirow{4}{*}{Truncation}
  & \multirow{2}{*}{Speaker} & 50\%  & $0.878$ & $0.814$ & $0.490$ \\
  &                          & 20\%  & $0.695$ & $0.663$ & $0.024$ \\
  \cmidrule{2-6}
  & \multirow{2}{*}{Record}  & 50\%  & $0.746$ & $0.679$ & $0.100$ \\
  &                          & 20\%  & $0.657$ & $0.616$ & $0.065$ \\
\bottomrule
\end{tabular}
\end{table}

Table~\ref{tab:perturb_defense} evaluates two inference-time perturbations applied to the reference audio before it is passed to the TTS model. This defense simulated possible defenses from model providers. The experiment setup is close to the perturbed query experiment. For noise addition, we add white noise at signal-to-noise ratios of 40 dB and 25 dB. A 40 dB SNR corresponds to barely perceptible noise and largely preserves perceived audio quality, while a 25 dB SNR causes clearly audible degradation but remains intelligible. These two levels cover a practical range that a deployed API might use without causing major quality concerns for regular users. Even under the stronger 25 dB noise setting, the attacks remain effective: speaker-level MIA still achieves a high AUC, and record-level MIA remains well above random. This suggests our proposed attacks are robust to substantial additive noise.

For temporal truncation, we keep only the first 50\% or 20\% of the original audio duration, corresponding to mild and aggressive truncation, respectively. Truncation reduces attack performance more strongly than additive noise, especially under the aggressive 20\% setting, where much of the reference speech is removed. However, the attacks still remain largely above random, indicating that membership signals can persist even when only a shortened reference is available. This also suggests that removing a large portion of the reference audio is necessary to substantially weaken the attack, which would likely reduce the utility of the TTS system for regular users.

\subsubsection{DP-SGD}
\label{sec:defense_dp}
\begin{table}[t]
\centering
\caption{Attack performance under DP-SGD defense for speaker-level and record-level MIA at two privacy budgets.}
\label{tab:dp_defense}
\setlength{\tabcolsep}{6pt}
\begin{tabular}{@{}llccc@{}}
\toprule
Level & $\varepsilon$ & AUC & ACC & TPR@1\%FPR \\
\midrule
Speaker & 4  & $0.533$ & $0.540$ & $0.008$ \\
        & 10 & $0.543$ & $0.555$ & $0.015$ \\
\midrule
Record  & 4  & $0.518$ & $0.515$ & $0.010$ \\
        & 10 & $0.534$ & $0.528$ & $0.006$ \\
\bottomrule
\end{tabular}
\end{table}

As is shown in Table~\ref{tab:dp_defense}, DP-SGD trained with privacy budgets $\varepsilon=4$ and $\varepsilon=10$ reduces both speaker-level and record-level MIA to near-random performance (AUC $\approx 0.52$--$0.54$, TPR@1\%FPR $<0.02$) on CosyVoice2 with VCTK, confirming that formal differential privacy effectively mitigates the membership leakage exposed by our attack.

\end{document}